%Paper: hep-th/9309110
%From: "Jacques Distler" <distler@puhep1.Princeton.EDU>
%Date: Mon, 20 Sep 93 18:11:29 EDT
%Date (revised): Fri, 5 Nov 93 15:51:41 EST

%% Print this file using Plain TeX
%  TeXing this paper requires harvmac.tex .
%  The figures require also epsf.tex .
%
\input harvmac
%%%%%%%%%%%%%%
% This inserts the figures using the macro package epsf.tex. If you don't have
% that macro package (available from hepth), or don't have the figure files,
% comment out this line:
%\def\figflag{I}
%%%%%%%%%%
% The following is the current definition for the macro \psheader, which issues
% the \special command to load a postscript header file. If the syntax for that
% \special command is different on your machine, change this line
% appropriately:
\def\psheader#1{}
%%
%%%%%%%%%%%

%%%%%%%%%%%%%%%%%%%%%%%%%%%%%%%%%%%%%%%%%%%%%%%%%%%%%%%%%%%%%%%%%%%%%%%%%%%
%Blackboard letters
%  The prehistoric version of this font is known as "msym". Many unfortunate
%  souls still have this old (and UGLY) ancestor of "msbm". Time to join the
%  modern world guys!

\font\blackboard=msbm10 \font\blackboards=msbm7
\font\blackboardss=msbm5
\newfam\black
\textfont\black=\blackboard
\scriptfont\black=\blackboards
\scriptscriptfont\black=\blackboardss
\def\blackb#1{{\fam\black\relax#1}}

%   Those truly poor slobs who have neither "msbm not "msym" fonts can
% substitute
%   the definition

%\def\blackb{\bf}

%   for the above font definitions or, if all else fails,
%   return to scratching symbols in the dirt with a sharpened stick.
%
\def\BC{{\blackb C}} 
 
\def\BZ{{\blackb Z}} 
\def\BP{{\blackb P}}

%%%%%%%%%%%%%%%%%%%%%%%%%%%%%%%%%%%%%%%%%%
% Math boldface letters
%
\font\mathbold=cmmib10 \font\mathbolds=cmmib7
\font\mathboldss=cmmib5
\newfam\mbold
\textfont\mbold=\mathbold
\scriptfont\mbold=\mathbolds
\scriptscriptfont\mbold=\mathboldss

%\def\bi{\bf}

%macros to get automatically-numbered figures
\def\tfig#1{Fig.~\the\figno\xdef#1{Fig.~\the\figno}\global\advance\figno by1}
%%%%%%%%%%
% Figure macros:
\def\figI{I}
%%%%%%%%%%%%%%%%:
%  This inputs the macro package epsf.tex and (on Unix)
%  loads the header for mac-produced figures
%
\newdimen\tempszb \newdimen\tempszc \newdimen\tempszd \newdimen\tempsze
\ifx\figflag\figI
\input epsf
%
% this complicated looking macro makes sure that a figure with given \epsfysize
%is narrow enough to fit inside the current \hsize. If not, it is scaled to
% fit.
% The ONLY reason this takes more than one line is that TeX doesn't do floating
% point arithmetic. So we have to go through some unexpected contortions....
\def\epsfsize#1#2{\expandafter\epsfxsize{
%calculate \tempsze = \epsfysize * \epsftsize / \epsfrsize
 \tempszb=#1 \tempszd=#2 \tempsze=\epsfxsize
     \tempszc=\tempszb \divide\tempszc\tempszd
     \tempsze=\epsfysize \multiply\tempsze\tempszc
     \multiply\tempszc\tempszd \advance\tempszb-\tempszc
     \tempszc=\epsfysize
     \loop \advance\tempszb\tempszb \divide\tempszc 2
     \ifnum\tempszc>0
        \ifnum\tempszb<\tempszd\else
           \advance\tempszb-\tempszd \advance\tempsze\tempszc \fi
     \repeat
%compare it to \hsize and do the right thing.
\ifnum\tempsze>\hsize\global\epsfxsize=\hsize\global\epsfysize=0pt\else\fi}}
\epsfverbosetrue
%\psheader{lprep68.procs} %this macro must be defined in the calling program.
\psheader{fig.pro}       %something like: \def\psheader#1{\special{header=#1}}
\fi
%
%%%%%%%%%%%%%%%%

%%%%%%%%%%%
%
% usage:  .... And now we gaze longingly at \tfig\figlabel. Later
%         we will have much to celebrate!
%
%         \ifigure\figlabel{caption}{figfile}{vsize}
%
%         Continuing in our petty pursuits...
%
% where vsize is the desired vertical size of the figure in truein
% If \figflag is undefined, then only a vbox is allocated; No
% \specials are called.

\def\ifigure#1#2#3#4{
\midinsert
\vbox to #4truein{\ifx\figflag\figI
\vfil\centerline{\epsfysize=#4truein\epsfbox{#3}}\fi}
\baselineskip=12pt
\narrower\narrower\noindent{\bf #1:} #2
\endinsert
}
%%%%%%%%%%%%%%%
%This one inserts 2 figures side by side:
%
%   \ifigures{\label1}{caption1}{file1}{vsize1}{\label2}{capt.2}{file2}{vsize2}
%
\def\ifigures#1#2#3#4#5#6#7#8{
\midinsert
\centerline{
\hbox{\vbox{
\divide\hsize by 2
\vbox to #4truein{\ifx\figflag\figI
\vfil\centerline{\epsfysize=#4truein\epsfbox{#3}}\fi}
\baselineskip=12pt
\narrower\narrower\noindent{\bf #1:} #2
}}\qquad
\hbox{\vbox{
\divide\hsize by 2
\vbox to #8truein{\ifx\figflag\figI
\vfil\centerline{\epsfysize=#8truein\epsfbox{#7}}\fi}
\baselineskip=12pt
\narrower\narrower\noindent{\bf #5:} #6
}}}
\endinsert
}

%%%%%%%%%%%%%%%%%%%%%%%%%%%%%%%%%%%%%%%%%%%%%%%%%%%%%%%%%%%%%%%%%%%%%

\def\appendix#1#2{\global\meqno=1\global\subsecno=0\xdef\secsym{\hbox{#1:}}
\bigbreak\bigskip\noindent{\bf Appendix #1: #2}\message{(#1: #2)}
\writetoca{Appendix {#1:} {#2}}\par\nobreak\medskip\nobreak}

% underlined text

\def\ket#1{| #1 \rangle}         % |#1>
         % <#1|
 %matrix element <#1|#2|#3>
%

 % del-bar
 % exterior power
\def\fourpt{\hbox{{$\rangle \kern-.25em \langle$}}} %cutesy Feynman diagram: ><
\def\tree{\hbox{{$\rangle \kern-.5em - \kern-.5em \langle$}}}%   :  >-<
 %dotless barred letters
 %

\def\H#1#2{{\rm H}^{#1}(#2)} %cohomology groups
 \def\CW{{\cal W}}

\def\ex#1{{\rm e}^{#1}}                 % exponential
\def\dd{\mskip 1.3mu{\rm d}\mskip .7mu} % exterior derivative

\def\cp#1{{\BC{\rm P}^{#1}}}
\def\wp#1{{W\BP^{#1}}}
\def\pd#1#2{{\partial #1\over\partial #2}}

\def\Ka{K\"ahler}
\def\cy{Calabi-Yau}
\def\LG{Landau-Ginzburg}
\def\sm{$\sigma$-model}
\noblackbox

\Title{\vbox{\hbox{PUPT--1419}\hbox{\tt hep-th@xxx/9309110}}}
{(0,2) \LG\ Theory$^\star$}

\centerline{Jacques Distler and Shamit Kachru}\smallskip
\centerline{Joseph Henry Laboratories}
\centerline{Princeton University}
\centerline{Princeton, NJ \ 08544 \ USA}
\bigskip\bigskip

\footnote{}{{\parindent=-10pt\par $\star$
\vtop{
\hbox{Email: {\tt distler@puhep1.princeton.edu}, and
{\tt kachru@puhep1.princeton.edu} .}
\hbox{Research supported by NSF grant PHY90-21984, and the
A.~P.~Sloan Foundation.}
     }     }}

A large class of (0,2) Calabi-Yau $\sigma$-models and
Landau-Ginzburg orbifolds
are shown
to arise as different ``phases'' of supersymmetric
gauge theories.
We find a phenomenon in the Landau-Ginzburg phase which may enable
one to understand which Calabi-Yau $\sigma$-models evade destabilization
by worldsheet instantons.  Examples of (0,2) Landau-Ginzburg vacua are
analyzed in detail, and several novel features of (0,2) models are
discussed.
In particular, we find that (0,2) models can
have different quantum symmetries from the (2,2) models built on the
same Calabi-Yau manifold, and that a new kind of topology change can
occur in (0,2) models of string theory.

%\draft                                    %%% later delete this
\Date{9/93}                 %%% and reinstate this
\lref\SpecialGeo{B. de Wit, P. Lauwers and A. van Pr\oe yen, Nucl. Phys. {\bf
B255} (1985) 569.}
\lref\ranspec{S. Cecotti, S. Ferrara and L. Girardello, Int. J. Mod. Phys.
{\bf A4} (1989) 2475\semi
L. Castellani, R. D'Auria and S. Ferrara, Class. Quantum Grav. {\bf 7} (1990)
1767.}
\lref\CandMod{P. Candelas and X. De la Ossa, Nucl. Phys. {\bf B355} (1991)
455.}
\lref\DKL{L. Dixon, V. Kaplunovsky and J. Louis, Nucl. Phys. {\bf B329}
(1990) 27.}
 \lref\StromSpec{A. Strominger, Comm. Math. Phys. {\bf 133}
(1990) 163.}
\lref\PerStrom{V. Periwal and A. Strominger, Phys. Lett. {\bf B335} (1990)
261.}
\lref\TopAntiTop{S. Cecotti and C. Vafa, Nucl. Phys. {\bf B367} (1991) 359.}
\lref\AandB{E. Witten, in ``Proceedings of the Conference on Mirror Symmetry",
MSRI (1991).}
\lref\LVW{W. Lerche, C. Vafa and N. Warner, Nucl. Phys. {\bf B324} (1989)
427.} %chiral ring
\lref\twisted{E. Witten. Comm. Math. Phys. {\bf 118} (1988) 411\semi
E. Witten, Nucl. Phys. {\bf B340} (1990) 281\semi
T. Eguchi and S.-K. Yang , Mod. Phys. Lett. {\bf A5} (1990) 1693.}%twisted N=2
 \lref\GVW{B. Greene, C. Vafa and N. Warner,
``Calabi-Yau Manifolds and Renormalization Group Flows,''
 {\it Nucl. Phys.} {\bf B324} (1989)
371.}
\lref\Grisaru{M. Grisaru, A. Van de Ven and D. Zanon, Phys. Lett. {\bf 173B}
(1986) 423.}
\lref\DSWW {M. Dine, N. Seiberg, X.G. Wen and E. Witten,
``Non-Perturbative Effects on the String World Sheet I,'' {\it Nucl.
Phys.}~{\bf B278} (1986) 769, ``Non-Perturbative Effects on the String
World Sheet II,'' {\it Nucl. Phys.}~{\bf B289} (1987) 319. }
\lref\DixonRev{L. Dixon, in ``Proceedings of the 1987 ICTP Summer Workshop in
High Energy Physics and Cosmology", ed G. Furlan, {\it et. al.}}
\lref\Exact{J. Distler and B. Greene, Nucl. Phys. {\bf B309} (1988) 295.}
\lref\twozero{J. Distler and B. Greene,
``Aspects of (2,0) String Compactifications,''
{\it Nucl. Phys.} {\bf B304} (1988) 1.}
\lref\AspMor{P. Aspinwall, and D. Morrison, ``Topological field theory and
rational curves", Oxford and Duke preprints OUTP-91-32P, DUK-M-91-12 (1991).}
\lref\CandMir{P. Candelas, X. de la Ossa, P. Green and L. Parkes,
``A Pair of Calabi-Yau Manifolds as an Exactly Soluble Superconformal
Theory,''
{\it Nucl.
Phys.} {\bf B359} (1991) 21.}
\lref\Kutconf{D. Kutasov, ``Geometry on the space of conformal field
theories and contact terms,'' Phys. Lett. {\bf B220} (1989) 153.}%
\lref\GrSei{M. Green and N. Seiberg, ``Contact interactions in
superstring theory," Nucl. Phys. {\bf B299} (1988) 559.}%
\lref\WilZee{F. Wilczek and A. Zee, Phys. Rev. Lett. {\bf 52} (1984) 2111.}
\lref\Banks{A. Sen, Nucl. Phys. {\bf B278} (1986) 289.\semi
T. Banks, L. Dixon,
D.Friedan and E. Martinec, Nucl. Phys. {\bf B299} (1988) 613.}
\lref\Hodge{P. Griffiths, Periods of Integrals on Algebraic manifolds I,II,
Am. J. Math. {\bf 90} (1970) 568,805\semi
R. Bryant and P. Griffiths, in ``Progress in Mathematics {\bf 36}"
(Birkh\"auser, 1983) 77.}
\lref\Odake{S. Odake, Mod. Phys. Lett. {\bf A4} (1989) 557\semi
S. Odake, Int. Jour. Mod. Phys. {\bf A5} (1990) 897\semi
T. Eguchi, H. Ooguri, A. Taormina and S-K. Yang, Nucl. Phys. {\bf B315}
(1989) 193.}
\lref\GSW{M. Green, J. Schwarz and E. Witten, ``Superstring theory, vol. II"
(Cambridge University Press, 1987).}
\lref\CHSW{P. Candelas, G. Horowitz, A. Strominger and E. Witten,
``Vacuum Configurations for Superstrings,'' {\it Nucl. Phys.} {\bf
B258} (1985) 46.}
\lref\NR{L. Alvarez-Gaum\'e, S. Coleman and P. Ginsparg, Comm. Math. Phys.
{\bf 103} (1986) 423.}
\lref\manin{D. Leites, ``Introduction to the theory of supermanifolds", Russ.
Math. Surveys {\bf 35} (1980) 3\semi
Yu. Manin, ``Gauge field theory and complex geometry", (Springer, 1988).}
\lref\Reviews{
J. Schwarz, ``Superconformal symmetry in string theory", lectures
at the 1988 Banff Summer Institute on Particles and Fields (1988)\semi
D. Gepner, ``Lectures on N=2 string theory",
lectures at the 1989 Trieste Spring School (1989)\semi
N. Warner, ``Lectures on N=2 superconformal theories and singularity theory",
lectures at the 1989 Trieste Spring School (1989)\semi
B. Greene, ``Lectures on string theory in four dimensions",
lectures at the 1990 Trieste Spring School (1990)\semi
S. Yau (editor), ``Essays in Mirror Manifolds",  Proceedings of the Conference
on Mirror Symmetry, MSRI (International Press, 1992).
}
\lref\Seiberg{N. Seiberg, Nucl. Phys. {\bf B303} (1988) 286.}
\lref\Dubrovin{B. Dubrovin, ``Geometry and integrability of
topological--antitopological fusion", INFN preprint, INFN-8-92-DSF (1992).}
\lref\GPM{B. Greene, D. Morrison, and R. Plesser, in preparation.}
\lref\GPmirror{B. Greene and R. Plesser, Nucl. Phys. {\bf B338} (1990) 15.}
\lref\phases{E. Witten, ``Phases of N=2 Theories in Two Dimensions",
{\it Nucl. Phys.} {\bf B403} (1993) 159, {\tt hep-th/9301042}.}
\lref\Vafa{C. Vafa, ``String Vacua and Orbifoldized LG models,''
{\it Mod. Phys. Lett.} {\bf A4} (1989) 1169.}
\lref\Ken{K. Intriligator and C. Vafa, ``Landau-Ginzburg Orbifolds,''
{\it Nucl. Phys.} {\bf B339} (1990) 95.}
\lref\Us{S. Kachru and E. Witten, ``Computing The Complete Massless
Spectrum Of A Landau-Ginzburg Orbifold,'' Princeton preprint PUPT-1397,
to appear in {\it Nucl. Phys.} {\bf B}, {\tt hep-th/9307038}.}
\lref\WitMin{E. Witten, ``On the Landau-Ginzburg Description of N=2
Minimal Models,'' IAS preprint
IASSNS-HEP-93/10, {\tt hep-th/9304026}.}
\lref\Fre{P. Fr\'e, F. Gliozzi, M. Monteiro, and A. Piras, ``A
Moduli-Dependent Lagrangian For (2,2) Theories On Calabi-Yau n-Folds,''
{\it Class. Quant. Grav.} {\bf 8} (1991) 1455; P. Fr\'e, L. Girardello,
A. Lerda, and P. Soriani, ``Topological First-Order Systems With
Landau-Ginzburg Interactions,'' {\it Nucl. Phys.} {\bf B387} (1992)
333, {\tt hep-th/9204041}.}
\lref\Greene{B.R. Greene, ``Superconformal Compactifications in Weighted
Projective Space,'' {\it Comm. Math. Phys.} {\bf 130} (1990) 335.}
\lref\fendley{P. Fendley and K. Intriligator, ``Central Charges
Without Finite Size Effects,'' Rutgers preprint RU-93-26, {\tt
hep-th/9307101}.}
\lref\OldWit{E. Witten, ``New Issues in Manifolds of SU(3) Holonomy,''
 {\it Nucl. Phys.} {\bf B268} (1986) 79.}
\lref\Pasquinu{S. Cecotti, L. Girardello, and A. Pasquinucci,
``Non-perturbative Aspects and Exact Results for the N=2 Landau-Ginzburg
Models,'' {\it Nucl. Phys.}~{\bf B338} (1989) 701, ``Singularity
Theory and N=2 Supersymmetry,'' {\it Int. J. Mod. Phys.} {\bf A6} (1991)
2427.}
\lref\KT{A. Klemm and S. Theisen, ``Mirror Maps and Instanton Sums for
Complete Intersections in Weighted Projective Space,'' Preprint LMU-TPW
93-08, {\tt hep-th/9304034}. }
\lref\VafaQ{C. Vafa, ``Quantum Symmetries of String Vacua,''
{\it Mod. Phys. Lett.}
{\bf A4} (1989) 1615. }
\lref\HWPmirrors{D. Morrison, ``Picard-Fuchs Equations and Mirror Maps for
Hypersurfaces," In {\it Essays on Mirror Manifolds}, ed. S.--T. Yau,
(Int. Press Co., 1992) {\tt alg-geom/9202026}\semi
A. Font, ``Periods and Duality Symmetries in Calabi-Yau
Compactifications,'' {\it Nucl. Phys.} {\bf B391} (1993) 358, {\tt
hep-th/9203084}\semi
A. Klemm and S. Theisen, ``Considerations of One Modulus Calabi-Yau
Compactification: Picard-Fuchs Equation, K\"ahler Potentials and Mirror
Maps,"
{\it Nucl. Phys.} {\bf B389} (1993) 153, {\tt hep-th/9205041}.}
\lref\flops{P. Aspinwall, B. Greene and D. Morrison, ``Multiple Mirror
Manifolds and Topology Change in String Theory," {\it Phys. Lett.} {\bf 303B}
(1993) 249, {\tt hep-th/9301043}.}
\lref\flopsII{P. Aspinwall, B. Greene and D. Morrison, ``Calabi-Yau Moduli
Space, Mirror Manifolds and Spacetime Topology Change in String Theory,"
IAS and Cornell preprints IASSNS-HEP-93/38, CLNS-93/1236, to appear.}
\lref\cvetic{M. Cveti\v c, ``Exact Construction of (0,2) Calabi-Yau
Manifolds,'' {\it Phys. Rev. Lett.} {\bf 59} (1987) 2829.}
\lref\Miron{J. Distler, B. Greene, K. Kirklin and P. Miron, ``Calculating
Endomorphism Valued Cohomology: singlet spectrum in superstring models,"
{\it Comm. Math. Phys.} {\bf 122} (1989) 117.}
\lref\miracles{M. Dine and N. Seiberg, ``Are (0,2) Models String Miracles?,"
{\it Nucl. Phys.} {\bf B306} (1988) 137.}

\newsec{Introduction}
The relationship between Calabi-Yau $\sigma$-models
with the gauge connection set equal to the spin connection
and (2,2) superconformal field
theories has been a rich source of conjectures and insights for
physicists (and mathematicians) in the past several years.  These
models are of particular interest to string theorists because they
provide compactifications of the heterotic string which lead to $N=1$
supersymmetric models of elementary particle physics \CHSW.

In the course of studying these phenomenologically promising string
vacua,
a wealth of ``stringy'' phenomena
which provide windows into how string theory modifies physics at
short distance scales
have been uncovered.
These string-theoretic modifications of our classical notions of
geometry are likely to provide important clues in uncovering a more
satisfactory formulation of the theory.

Of course, from the point of view of trying to understand the space of
ground states of
classical string theory, or even trying to find
phenomenologically promising vacua, there is
little reason to restrict oneself to the study of (2,2) theories.
Indeed, (0,2) Calabi-Yau theories were introduced {\it precisely}
because they offer improved phenomenological prospects, compared to
their (2,2) counterparts \OldWit.  In particular, while (2,2) theories
lead to an observable $E_{6}$ unification group in spacetime, (0,2)
theories with $SO(10)$ or $SU(5)$ gauge group can be constructed by
choosing the vacuum gauge bundle to have rank four or five.

Beyond a few preliminary explorations of (0,2) Calabi-Yau models and
orbifolds \refs{\twozero,\cvetic,\Greene}, however, little is known about
(0,2) theories.  This is largely because of the assertion in \DSWW\ that
the generic (0,2) \sm\
is destabilized by worldsheet instantons,
topologically non-trivial maps from the string worldsheet to rational
curves in the Calabi-Yau target space.
Although it was pointed out in \twozero\ that vacuum gauge bundles which
{\it split} non-trivially over all of the rational curves in the
Calabi-Yau manifold would evade this destabilization, explicit
verification of that condition in non-trivial
examples is prohibitively difficult. Therefore, without any classes of
explicit and tractable (0,2) vacua to study, this
vast region of the moduli space of classical string vacua has remained
largely unexplored.

Recently, Witten pointed out \phases\ that
one can quite explicitly construct
certain massive (0,2) $\sigma$-models which
are (0,2) deformations of the more
conventional (2,2) Landau-Ginzburg orbifolds \Vafa; he further argues
that there are no instantons which could destabilize these (0,2)
Landau-Ginzburg theories.
As in the (2,2) case,
(0,2) Landau-Ginzburg orbifolds are continuously connected to
Calabi-Yau models in a non-singular ``phase diagram,'' so
it seems likely that by studying Landau-Ginzburg theory, one can
learn about Calabi-Yau vacua again in the (0,2) case.
Because
(0,2) Landau-Ginzburg theories can be constructed quite explicitly using
the methods of \Us,  one finally has a large class of explicit
(0,2) vacua available for detailed examination.

In the present work, we undertake the
detailed study of (0,2) Landau-Ginzburg vacua, including (0,2) models
which are not merely deformations of (2,2) theories.  Our results
indicate that the study of (0,2) string vacua is likely to reveal many
surprises unanticipated from the (2,2) case.

In \S2, we proceed in the style of \phases\ to examine the full phase
diagram of the massive theories which flow, in the infrared, to a
class of (0,2) Calabi-Yau $\sigma$-models much more general than that
discussed in \phases.
We find several surprises in the (0,2) case; for
example, (0,2) models  built on complete intersection Calabi-Yau manifolds
admit a Landau-Ginzburg representation, while the corresponding (2,2) models
did not.

In \S3, we discuss the Landau-Ginzburg phase of generic (0,2) theories
in more detail, generalizing the results of \Us.  We find that the
Landau-Ginzburg theories provide a natural and physical criterion which
might indicate which (0,2) theories are destabilized by worldsheet
instantons and which are not.
Some (0,2) models seem to exist only in highly restricted
subspaces of their purported moduli space, while others
(including non-deformation (0,2) theories with phenomenologically
interesting gauge groups)
appear to make
sense generically.

In \S4, we present an example of each type
of (0,2) model.
First, we present an
exhaustive analysis of a
complete intersection model with $SO(10)$ gauge
group.  The Landau-Ginzburg phase seems to be sensible for generic
choices of its defining data.  We then present another example which
cannot be  sensibly embedded in a heterotic string theory for generic choices
of its defining data; for non-generic choices of a special type, however, the
model still makes sense, and defines a consistent (0,2) theory.  This
model is therefore constrained to lie on a small subspace of its naive moduli
space. Presumably, this is the Landau-Ginzburg symptom of the worldsheet
instanton disease found in the Calabi-Yau phase.

\S5 contains a discussion of some very interesting novel features
exhibited by the (0,2) vacua we have studied.  In particular,
we find that such vacua exhibit new quantum symmetries, distinct from their
(2,2) cousins defined in the same $\wp{n}$, and we provide an example
which suggests that a new kind of topology change occurs in (0,2)
models.

\newsec{Microscopic Lagrangians, and the Phase Structure of (0,2) Models}

To obtain a heterotic string theory in 4 dimensions with SO(10) or SU(5) gauge
group, we need as the ``internal" part of the theory a (0,2)
superconformal field theory with central
charge  $(c,\bar c)=(6+\tilde r,9)$, where $\tilde r$ is 4 for $SO(10)$,
and 5 for $SU(5)$. In addition to the right-moving $N=2$ algebra, there
should also be a left-moving $U(1)$ current algebra with
$$J(z)J(w)={\tilde r\over (z-w)^2}$$
 so that the linearly
realized part of the gauge group is $SO(8)\times U(1)$ or $SO(6)\times U(1)$,
respectively. This left-moving $U(1)$ is used to implement the GSO projection
for the left-moving (``gauge") fermions, just as $\bar J$, the $U(1)$ in the
right-moving $N=2$ algebra is used to implement the GSO projection on the
right-moving (``Lorentz") fermions. By the analogue of spectral flow, the
representations of the linearly-realized part of the gauge group assemble
themselves into representations of the full gauge group (table (1)).

The most familiar example of such an SCFT is a (0,2) \sm\ on a
\cy\ manifold $M$. The ``new" data in a (0,2) \sm\ is a choice of a rank
$\tilde r$ stable holomorphic vector bundle $E\to M$, with $c_1(E)=0$ and
$c_2(E)=c_2(T)$, where $T$ is the holomorphic tangent bundle of $M$
\OldWit.

\def\tablerule{\omit&\multispan{6}{\tabskip=0pt\hrulefill}&\cr}
\def\tablepad{\omit&height3pt&&&&&&&\cr}
$$\vbox{\offinterlineskip\tabskip=0pt\halign{
\strut$#$\quad&\vrule#&\quad\hfil $#$ \hfil\quad &\vrule #&\quad \hfil $#$
\hfil \quad&\vrule #& \quad $#$ \hfil\ &\vrule#&\quad $#$\cr
&\omit&\hbox{Rep.~of $SO(10)$}&\omit&\hbox{Rep.~of $SO(8)\times
U(1)$}&\omit&\hbox{Cohomology Group}&\omit&\cr
\tablerule\tablepad
&&45&&8^{s'}_{-2}\oplus28_0\oplus1_0\oplus 8^{s'}_2&&\H{*}{M,\CO}&&\cr
\tablerule\tablepad
\tilde r=4&&16&&8^{s}_{-1}\oplus8^{v}_1&&\H{*}{M,E}&&\cr
\tablerule\tablepad
&&10&&1_{-2}\oplus8^{s'}_{0}\oplus1_2&&\H{*}{M,\bigwedge^2E}&&\cr
\tablerule\tablepad
&&1&&1_0&&\H{*}{M,End\ E}&&\cr \tablerule
\noalign{\bigskip}
&\omit&\hbox{Rep.~of $SU(5)$}&\omit&\hbox{Rep.~of $SO(6)\times
U(1)$}&\omit&\hbox{Cohomology Group}&\omit&\cr
\tablerule\tablepad
&&24&&\bar 4_{-5/2}\oplus15_0\oplus1_0\oplus 4_{5/2}&&\H{*}{M,\CO}&&\cr
\tablerule\tablepad
\tilde r=5&&10&&4_{-3/2}\oplus6_1&&\H{*}{M,E}&&\cr
\tablerule\tablepad
&&\bar 5&&\bar 4_{-1/2}\oplus 1_2&&\H{*}{M,\bigwedge^2E}&&\cr
\tablerule\tablepad
&&1&&1_0&&\H{*}{M,End\ E}&&\cr \tablerule
\noalign{\bigskip}
\noalign{\narrower\noindent{\bf Table 1:} Representations of the
linearly realized part of the gauge group and how they assemble themselves.}
 }}$$

In a very beautiful paper \phases, Witten showed how one can
construct a (2,2) \cy\ \sm\ as the infrared limit of a
(2,2) supersymmetric quantum field theory. Moreover, as one varies
the parameters of this ``microscopic" theory, the theory exhibits
different phases. In the simplest case, there are two phases.
One is the
\cy\ \sm; the other is a (deformed) \LG\ theory.

We would like to follow in the same spirit and construct the
microscopic (0,2) theory whose IR limit is the desired (0,2) \cy\ \sm, and
then examine its phase structure.

The first step in this construction is to specify the stable vector bundle
$E\to M$ which couples to the left-moving worldsheet fermions.
In the (2,2) case, this was just $T$, the tangent bundle of $M$.
One method
of defining $E$, which has a simple field-theoretic realization, is the
following. For simplicity, we will consider the simple case of a hypersurface
in $\wp{4}$, with
homogeneous coordinates $\phi_i$. Consider the exact sequence
\eqn\efundseq{0\to E\to \bigoplus_{a=1}^{\tilde r+1} \CO(n_a)  {\buildrel
{\otimes F_a(\phi)}\over{\hbox to 30pt{\rightarrowfill}}}  \CO(m)\to 0} where
the $n_a$ are positive integers\foot{The $n_a$ must be strictly positive,
otherwise $E$ is {\it never} stable.},  $m=\sum n_a$, and the $F_a$ are
homogeneous polynomials of the appropriate degrees. We demand that the $F_a$ be
such that at no point in $M$ do they all vanish simultaneously. If the $F_a$
are chosen sufficiently generically, then the kernel $E$ is a
stable rank $\tilde r$
vector bundle with vanishing first Chern class, $c_1(E)=0$. The ``Maruyama
construction" (of the dual bundle $E^*$) discussed in  \twozero\ is a special
case of the above, in which $\tilde r$ of the $n_a$ are chosen to be equal to
1.

If the $F_a$ are chosen ``badly", then $E$ is in general
a semistable torsion-free sheaf. However, just as ``bad" choices of defining
polynomial can lead to a singular variety, which nonetheless is perfectly
fine as a string theory, here too the string theory that results from
a nongeneric choice of $F_a$'s is well-behaved.

As in the (2,2) case \phases, we introduce (our (0,2) conventions are found in
the appendix) a set of chiral superfields
$\Phi_i$ with $U(1)$ gauge charges  equal to $w_i$,
the weights of the corresponding homogeneous coordinates in $\wp{4}$. The
hypersurface is the zero locus of a degree $d$ polynomial $W(\phi)$  where the
\cy\ condition ({\it i.e.}~$c_1(T)=0$) is $d=\sum w_i$. To enforce $W=0$, we
will need  a Fermi superfield $\Sigma$
of charge $-d$. To construct the bundle $E$ above, we introduce some Fermi
superfields $\Lambda^a$ with charges  $n_a$, and a chiral superfield $P$ with
charge $-m$. Finally, we have the (0,2) gauge multiplets, $V$ and $\CA$. In
Wess-Zumino gauge, $\CA$ contains the left-moving gauge field $a$, a pair
of left-moving gauginos $\alpha,\tilde\alpha$, and the auxiliary field $D$.
$V$ contains the right-moving component of the gauge field $\bar a$. The (0,2)
superpotential is
\eqn\esuperpotI{S_\CW=\int \dd^2 z\dd\theta\ \Sigma W(\Phi) + P\Lambda^a
F_a(\Phi)} The first term will ensure that we lie on the hypersurface
$W(\phi)=0$, and the second term will insure that the $\lambda^a$ (the lowest
components of the $\Lambda^a$) are sections of the bundle $E$.

Of course, the bundle $E$ must satisfy a constraint on its second Chern class:
\eqn\ectwo{c_2(E)=c_2(T)}
In the present context, this is the equation
\eqn\ectwonow{m^2-\sum n_a^2=d^2-\sum w_i^2}
which is none other than the condition for the cancellation of the U(1) gauge
anomaly for the fields in our theory.

This is a Diophantine equation for the
positive integers $n_a$, and in general, we won't find many solutions. We are,
however, always guaranteed at least one.
Let $\tilde r=4$, and set $n_i=w_i$, then
the equation is satisfied identically. This solution for $E$ is simply a stable
deformation of $T\oplus\CO$ (more precisely, of an extension of $T$ by $\CO$).
Indeed, $E$ is isomorphic to such an extension precisely when
$F_i(\phi)=\pd{W}{\phi_i}$.

To see this, we need only
examine the definition of the tangent bundle $T$ of a \cy\ hypersurface.
Consider the sequence
\eqn\eTseq{0\to
\CO
{\buildrel u\over\longrightarrow}
%{\buildrel\otimes \phi_i          \over{\hbox to 20pt{\rightarrowfill}}}
\bigoplus_{i=1}^{5} \CO(w_i)
{\buildrel v\over\longrightarrow}
%{\buildrel\otimes\partial_i W(\phi)\over{\hbox to 30pt{\rightarrowfill}}}
\CO(d)\to 0}
where
$$u(f)=(w_{1}\phi_1f,w_{2}\phi_2f,\dots,w_{5}\phi_5f),
\qquad v(s_1,s_2,\dots,s_5)=\sum_i s_i\pd{W}{\phi_i}$$
 This sequence is clearly not exact (the dimensions don't
add up), but nonetheless, the image of $u$ is contained in the
kernel of $v$, since we have $\sum_i w_i\phi_i \partial_iW(\phi)= d\
W(\phi)=0$ on $M$. The {\it cohomology} of this sequence, ${\rm ker}(v)/{\rm
im}(u)$ is the tangent bundle $T$. For this very special defining
data, we thus have an exact sequence $0\to\CO\to{\rm ker}(v)\to
T\to0$.\foot{Although we have only shown that
${\rm ker}(v)$ is an extension of $T$ by $\CO$, one can show using the methods
of
\S3 that the physical
spectrum of the  corresponding \LG\ theory is isomorphic to that based on
$T\oplus\CO$ ({\it i.e.}~the corresponding (2,2) theory). The point is that
the
$\bar Q_{+}$ operator differs from that of the $T\oplus\CO$ theory simply by
the
presence of a
$\sigma W(\phi)$ term in
$\bar Q_{+,L}$. However, when $F_i=\pd{W}{\phi_i}$, this term is  $\bar
Q_{+,L}$ trivial. Thus the $\bar Q_{+}$ cohomology is the same as when this
term is absent, {\it i.e.}~isomorphic to that of the (2,2) theory.} The
definition of
$E$ is simply to replace
$\partial_i W$ in the definition of the map $v$ by generic polynomials $F_i$
of the same degree, so that, for generic $F_i$, $E$ is stable.
 Thus the (0,2) theory discussed in \twozero\ and many of the (0,2) theories in
\Greene\ are simply deformations of the corresponding (2,2) theories.

The net number of generations of matter fields (16s of $SO(10)$ or $(\bar 5
+10)$s of $SU(5)$) is $\half c_3(E)$. If $M$ is smooth (which is all we
will consider in this paper), $c_3(E)$ is given by
\eqn\ecthree{c_3(E)={1\over 3}(m^3-\sum_a n_a^3)J^3\quad .}
For these deformations of
$T\oplus\CO$, the number of generations is, of course, exactly the same as
that of the corresponding (2,2) model. For this reason, we will mostly concern
ourselves with (0,2) theories which {\it are not} deformations of (2,2)
theories.

Integrating out the the $D$ auxiliary field in the gauge multiplet and the
auxiliary fields in the Fermi multiplets, we obtain the scalar potential
\eqn\escpot{U=|W(\phi)|^2 +|p|^2|F_a(\phi)|^2 +{e^2\over2}\left( \sum_i
w_i|\phi_i|^2\ - m|p|^2-r\right)^2}
where $r$ is the coefficient of the Fayet-Iliopoulos D-term.

In general $r$ gets additively renormalized at one loop \phases. The
counterterm is proportional to the sum of the $U(1)$ charges of the scalar
fields in the theory
\eqn\edeltar{\Delta r_{\rm 1-loop}\propto \sum_{\rm bosons} Q_i}
For the model we have been discussing, this sum is equal to $(d-m)$. In
general, this might not be zero, in which case, $r$ runs as a function of
RG scale. This is inconvenient, since we were hoping to use $r$ as an
RG-invariant parameter to label the infrared fixed points that we obtain. In
particular, $t=r+i\theta$ was to parametrize the \Ka\ moduli space of the
model.

The situation, however, is easily saved. Simply introduce a chiral superfield
$X$ of charge $(m-d)$, and a Fermi superfield $\Omega$ of charge $(d-m)$ and
add the term
\eqn\espectsuppot{\Delta S_\CW=\int \dd^2 z\dd\theta\ \Omega X}
to the superpotential. The addition of these new ``spectator" fields cancels
the 1-loop contribution to the $D$ term, restoring $t$ to its status as an RG
invariant.

The scalar potential is, of course modified by the addition of the
scalar $x$. However, minimizing $U$ {\it always} uniquely fixes $x=0$, and the
spectators, being massive, drop out of the low-energy physics.  For the most
part, we will simply ignore the effects of the spectators, save for their
making $r$ an RG-invariant.

\subsec{The \cy\ phase}

For $r\gg0$, the semiclassical approximation is a good guide to determine
the ground state of the theory. To find the ground state, we simply need to
minimize the classical scalar potential.  The minimum is obtained by
setting $p=0$, $\sum w_i|\phi_i|^2=r$, and $W(\phi)=0$. This spontaneously
breaks the gauge symmetry and one component of $\phi$ gets eaten by the gauge
field. Three more pick up masses from \escpot. The remaining three (complex)
massless $\phi$'s parametrize the \cy\ hypersurface $M$.

The right-moving fermion $\pi$, the superpartner of $p$, picks up a mass with
one linear combination of the left-moving fermions $\lambda^a$, through the
Yukawa coupling
$$\pi\lambda^a F_a(\phi)$$
which follows from the superpotential \esuperpotI. The remaining massless
$\lambda^a$'s transform as local sections of the bundle $E$ defined by
\efundseq.
The left-moving fermion $\sigma$ in $\Sigma$ picks up a mass with one linear
combination of the right-moving fermions $\psi_i$ in $\Phi_i$ through the
Yukawa coupling
$$\sigma \psi_i\pd{W}{\phi_i}$$
The remaining $\psi$'s are thus in the kernel of the map $v$ in \eTseq.
The Yukawa coupling with the gaugino
\eqn\egaugeyuk{-\sum_i w_i\alpha\psi_i\bar\phi_i}
gives mass to another linear
combination of the $\psi$'s, getting rid of the image of the map $u$ in \eTseq.
The
remaining massless right-moving fermions thus transform as local sections of
the tangent bundle $T$. Thus the massless fields form a (0,2) \cy\ \sm\ in
which the scalars $\phi$ parametrize a \cy\ hypersurface $M$, the right-moving
fermions $\psi$  are sections of the tangent bundle $T$, and the left-moving
fermions $\lambda$ are sections of the bundle $E$.

\subsec{The \LG\ phase}

For $r\ll0$, we again have to minimize the scalar potential. Now the minimum
is obtained for $|p|^2=|r|/m$, $\phi=0$. The gauge symmetry is again
spontaneously broken. Again, $p$ becomes massive and drops out of the low
energy theory, as does its superpartner $\pi$, this time through its
Yukawa coupling with the  gaugino
$$m\alpha \pi\bar p$$
What remains is a \LG\ theory with the superpotential
(after a trivial rescaling of the fields)
\eqn\esuperpotLG{S_\CW=\int \dd^2 z\dd\theta\ \Sigma W(\Phi)+ \Lambda^a
F_a(\phi)}
Actually, since the field $p$ which got the VEV has charge $m$, a $\BZ_m$
subgroup of the gauge group remains unbroken, so we, in fact, have a \LG\
{\it orbifold}.  This will be the subject of study in \S3.

\subsec{The point $r=\theta=0$}

At $r=0$, the classical potential is minimized for $p=\phi^i=0$,
corresponding to unbroken gauge symmetry. In the (2,2) case, the analysis of
the behaviour of the theory in the neighbourhood of this point is quite
delicate. At $r=0$, the classical potential for the scalar superpartner of the
gauge field
(called $\sigma$ in \phases) becomes flat, and semiclassical reasoning becomes
invalid. A somewhat delicate analysis (involving working at nonzero $\theta$)
is required to show that spectrum of low-lying states (when the theory is
quantized on a circle) is discrete, and varies
continuously as one passes through $r=0$ \phases.

In our (0,2) theories, showing this is a triviality. There is no scalar
superpartner of the gauge field whose spectrum might become continuous. All of
the scalar fields have potentials that grow at infinity, so the low-lying
spectrum is always discrete, even at $r=0$.

\subsec{Generalizations}

There are {\it lots} of generalizations of the basic construction we have
outlined. Let us look at a few of them.

The first obvious generalization is to
look at \cy\ manifolds realized as complete intersections in weighted
projective space. Introduce several Fermi superfields $\Sigma_j$, with
charges $-d_j$, where $d_j$ are the degrees of the polynomials
$W_j(\phi)$. Instead of \esuperpotI, consider the more general superpotential
\eqn\esuperpotCI{S_\CW=\int \dd^2 z\dd\theta\ \Sigma_j W_j(\Phi) + P\Lambda^a
F_a(\Phi)} The \cy\ condition is now $\sum d_j=\sum w_i$, and the
condition that $c_2(E)=c_2(T)$ now reads
\eqn\ectwoCI{m^2-\sum n_a^2=\sum d_j^2-\sum w_i^2}
Again, this is just the condition that the gauge symmetry be nonanomalous.
As before, for $r\gg0$, we clearly have a (0,2) \cy\ \sm\ on the \cy\ manifold
which is the simultaneous vanishing locus of all the $W_j$'s.

Our first surprise comes when we look at $r\ll0$. In the corresponding (2,2)
case, this is a ``hybrid phase" \phases\ about which little concrete can at
present be said. Here, however, $r\ll0$ is again simply a (0,2) \LG\ orbifold.
The point is that, whereas in the (2,2) theory, there are several
negatively-charged scalars which could in principle pick up a VEV, here there
is still only one -- $p$ -- and so we have a \LG\ phase, rather than a hybrid.
This is but the first hint that the \Ka\ moduli space of (0,2) theories is in
general very different from that of the corresponding (2,2) theory. We will see
this more explicitly in \S5.

We can also consider complete intersections in products of
weighted projective spaces. Introduce two $U(1)$ gauge groups and two sets of
chiral multiplets: $\Phi_i$ with charges $(w_i,0)$ and $\Upsilon_i$ with
charges $(0,\tilde w_i)$. Then let the superpotential be
\eqn\esuperpotPP{S_\CW=\int \dd^2 z\dd\theta\ \Sigma_j W_j(\Phi,\Upsilon) +
P\Lambda^a F_a(\Phi,\Upsilon)}
Again, the condition $c_2(E)=c_2(T)$ is simply the requirement that the gauge
anomalies vanish. There are now two Fayet-Iliopoulos D-terms and,
correspondingly, four phases depending on the signs of the
coefficients of these D-terms. One phase, when both of the $r_i$ are
positive, is the familiar \cy\ phase. When both of the $r_i$ are negative, it
is $p$ that develops an expectation value. This leaves a $U(1)$ subgroup of
the gauge group unbroken. Thus this phase is a gauged \LG\ theory.
 The remaining phases are hybrids.

There is one more generalization that has no analogue in the (2,2) case. We
can have two (or more) factors in the vacuum gauge bundle $E$, $E=E_1\oplus
E_2$, and if we wish, we can embed $E_1$ in the first $E_8$ and $E_2$ in the
second $E_8$ of the $E_8\times E_8$ heterotic string. To represent this
situation, we can generalize \esuperpotPP\ still further:
\eqn\esuperpotPPMG{S_\CW=\int \dd^2 z\dd\theta\ \Sigma_j W_j(\Phi,\Upsilon) +
P_1\Lambda^a_1 F_a(\Phi,\Upsilon)+P_2\Lambda^a_2 G_a(\Phi,\Upsilon)}
where the $F_a$ are the polynomials which define the bundle  $E_1$ and $G_a$
are the polynomials which define $E_2$.
 A complete intersection in a product of
{\it two} weighted projective spaces, with {\it two} factors in the vacuum
gauge bundle again has a \LG\ phase.

Finally, we can consider generalizations of our basic construction of the
vector bundle $E$. For instance, we can generalize \efundseq\ to
\eqn\egenseq{0\to E\to\bigoplus_{a=1}^{\tilde r+2}\CO(n_a)\to
\CO(m_1)\oplus\CO(m_2)\to 0}
This is accomplished by generalizing \esuperpotI\ to
\eqn\esuperpotICI{S_\CW=\int \dd^2 z\dd\theta\ \Sigma W(\Phi) + P_1\Lambda^a
F_a(\Phi)+ P_2\Lambda^a
\tilde F_a(\Phi)}
where $P_1$ has charge $-m_1$ and $P_2$ has charge $-m_2$.

Clearly, the variations on this theme are nearly endless. However, aside from
being able to write down the expected phase diagram, one cannot, in most of
these models, say much more. In particular, one would like to show that the
models do indeed {\it have} (0,2) SCFT's as their IR limits. In the context of
\cy\ \sm s, doubt has been cast on the existence of the infrared fixed point
due to worldsheet instanton effects \DSWW.

In order to bolster our confidence that at least a subclass of the microscopic
quantum field theories we have been studying really do have (0,2) SCFT's
as their IR limits, we will focus on those that have a \LG\ phase. The \LG\
theory {\it can} be constructed quite explicitly using the methods of
\Us.\foot{Very similar techniques were first developed in the context of
(2,2) theories in \refs{\Vafa,\Ken,\Pasquinu}.}
At least for these theories, we can know with confidence that the LG phase
has as its IR limit a (0,2) SCFT. Since at nonzero $\theta$, the LG phase
interpolates smoothly into the \cy\ phase (the only singularity is at the point
$r=\theta=0$), we can be confident that the \cy\ phase exists as well. In
addition, we will be able to compute the massless spectrum of the LG theory,
and compare the results with those computed in the \sm.

\newsec{The \LG\ Phase}

Consider the Landau-Ginzburg theory with superpotential
\eqn\sup{{\cal W}(\Phi_{i},\Lambda^{a},\Sigma_{j}) =
\Lambda^{a}F_{a}(\Phi_{i}) +
\Sigma^{j}W_{j}(\Phi_{i})}
where $a=1,\dots,{\tilde r}+1$, $i=1,\dots,N+1$, $j=1,\dots,N-3$.
This is the effective superpotential that remains in the \LG\ phase of
\esuperpotCI, after setting $P$ equal to its vacuum expectation value.
For simplicity, we have not included the ``spectator multiplets'' $\Omega, X$,
though we clearly could do so with little extra effort. Including them changes
none of the subsequent analysis. Their effect is solely to cancel the
renormalization of the Fayet-Iliopoulos D-term.

Transversality of the defining equations $W_{j}(\phi) = 0$ and
our conditions on the $F_a(\phi)$ which define the bundle $E$ in the Calabi-Yau
phase are sufficient
to insure than in the Landau-Ginzburg phase, the superpotential \sup\
is a
quasi-homogeneous function of the
fields with an isolated singularity at the origin of field space.
In particular, the $U(1)$ gauge charges
$w_{i}, n_{a},-d_{j}$ and $m$ are integers which satisfy
\eqn\quasi{{\cal W}(\epsilon^{w_{i}}\Phi_{i},
\epsilon^{n_{a}}\Lambda^{a},\epsilon^{-d_{j}+m}\Sigma_{j})
 = \epsilon^{m}{\cal W}(\phi_{i}, \Lambda^{a}, \Sigma^{j})~.}
The quasi-homogeneity \quasi\ of the superpotential guarantees the
existence of a right-moving $U(1)$
R-symmetry which plays an important role in (0,2) models.

Actually, the models we are interested in have  an additional global
left-moving $U(1)$ symmetry.
Let us define
\eqn\defq{q_{i}={w_{i}\over m}, ~~q_{a}={n_{a}\over m},~~q_{j}=
{1-{d_{j}\over m}}~.}
Then the left and right $U(1)$ charges of the various fields are given
in table (2).

\def\tablerule{\omit&\multispan{7}{\tabskip=0pt\hrulefill}&\cr}
\def\tablepad{\omit&height3pt&&&&&&&\cr}
$$\vbox{\offinterlineskip\tabskip=0pt\halign{
\hskip.5in\strut$#$\quad&\vrule#&\quad\hfil $#$ \hfil\quad &\vrule #&\quad
\hfil $#$
\hfil \quad&\vrule #& \quad\hfil $#$ \hfil \quad&\vrule#&\quad $#$\cr
&\omit&\hbox{Field}&\omit&\hbox{Left $U(1)$ charge $q$
}&\omit&\hbox{Right $U(1)$ charge $\bar q$}&\omit&\cr
\tablerule\tablepad
&& \phi_{i} &&q_{i}&&q_{i}&&\cr
\tablerule\tablepad
&&\psi^{i}&&q_{i}&&q_{i}-1&&\cr
\tablerule\tablepad
&&\lambda^{a}&&q_{a}-1&&q_{a}&&\cr
\tablerule\tablepad
&&\sigma^{j}&&q_{j}-1&&q_{j}&&\cr
\tablerule
\noalign{\bigskip}
\noalign{\narrower\noindent{\bf Table 2:}
Left and right $U(1)$ charges of fields in Landau-Ginzburg theory.}
 }}$$

\bigskip

We are only interested in describing Landau-Ginzburg theories which
are the $r \rightarrow -\infty$ limit of some ``high energy theory''
with a Calabi-Yau description at $r\rightarrow \infty$.
Hence, as in \S2, we require that the charges satisfy various quadratic
conditions so that the gauge and global $U(1)$ symmetries are
non-anomalous, and of course, we impose
\eqn\ecy{\sum_{i} w_{i} =  \sum_{j} d_{j}}
so that at $r\rightarrow \infty$,
the equations $W_{j}(\phi_{i})=0$ describe a
Calabi-Yau manifold in
$\wp{N}$.

For string theory, we are interested in the {\it conformal} limit of the
theory defined by \sup. It is, as we have emphasized, only an {\it
assumption} that the quantum field theory we are discussing has as its IR
limit a (0,2) SCFT (the same might be said about the (2,2) quantum
field theories in \phases). Nevertheless, {\it making this assumption}, we
will be able to say quite a lot about the properties of that SCFT.
That is the goal of this section. In \S3.3, we will see that there are some
nontrivial conditions to be satisfied for this SCFT to be embedded in a
string theory.

  Because the left-moving stress-tensor $T(z)$ is an element
of the chiral algebra formed by the $\bar Q_{+}$ cohomology  (see \WitMin), the
conformal anomaly of the left-moving CFT is reliably computed in the massive
theory and is given by
\eqn\leftc{c = \sum_{a} (1-3 { q_{a}}^{2})
+ \sum_{j} (1-3 { q_{j}}^{2}) + \sum_{i} (2+3 {
q_{i}}^{2} - 6 q_{i})~.}
Using the quadratic condition
\eqn\quadratic{\sum_{a} q_{a}^{2}
+ \sum_{j} q_{j}^{2} = \sum_{i}
q_{i}^{2} + 1}
and the condition \ecy, a few steps of algebra reveal that
for the theory \sup\ the left central charge \leftc\ simplifies to
\eqn\leftagain{c = 6 + {\tilde r}}
as expected.

Similarly, the left-moving $U(1)$ current, $J(z)$,  is in the
chiral algebra formed by the $\bar Q_{+}$ cohomology, and so we
can accurately compute, in the massive theory, the central term in the $J\cdot
J$ operator product expansion.
It is given by
\eqn\ekleft{\sum_a (q_a-1)^2+\sum_j (q_j-1)^2-\sum_i q_i^2=\tilde r}
as expected.

In the standard way, we supplement the Landau-Ginzburg
theory with $16-2{\tilde r}$ extra left-moving free fermions
\eqn\extra{L' = \int \sum_{I=1}^{16-2{\tilde r}}
\lambda^{I}\bar\partial \lambda^{I}}
which generate an $SO(16-2{\tilde r})$ current algebra
(which combines with the
left-moving $U(1)$ of the Landau-Ginzburg theory to yield a maximal
subgroup of the visible spacetime
gauge group, which is $E_{6}$ for ${\tilde r}=3$,
 $SO(10)$ for ${\tilde r}=4$, and $SU(5)$
for ${\tilde r}=5$).  There are
also 16 left-moving free fermions which generate
the ``hidden'' $E_{8}$ spacetime gauge symmetry; we shall leave them in
their NS sector for the duration of this paper, and only consider states
which are singlets under the ``hidden'' gauge group.

What about the right-moving central charge?
We cannot reliably compute
$\bar c$ from the $\bar T(\bar z)\bar T(0)$
operator product in the massive theory.  However, $\bar c$ appears in
other places in the N=2 superconformal algebra; in particular,
the central term in the $\bar J\cdot\bar J$ operator product is $\bar
c/3$. This is nothing other than the anomaly (of a fictitious gauge field
coupled to $\bar J$). This can be computed reliably by a one-loop
computation in the massive theory\foot{This trick, of using the $U(1)$ anomaly
to compute the $N=2$ central charge, has been exploited by Fendley and
Intriligator in a somewhat different context for perturbed $N=2$ theories
\fendley.}. The result is as expected:
\eqn\rightc{{{\bar c}\over 3}  =  \sum_{i}(q_{i}-1)^{2} -
\sum_{a}
 q_{a}^{2} = 3}
which again follows from using the conditions \ecy\ and \quadratic.

\subsec{Twisted Sectors and Quantum Numbers}

Recall that
in the passage from the high energy theory with superpotential \esuperpotCI\ to
the low energy theory with superpotential \sup\ at $r \ll0$,
the $P$ field develops an expectation value
\eqn\pexp{|P|=\sqrt{-r\over m}}
and spontaneously breaks the $U(1)$ gauge symmetry to a residual
$\BZ_{m}$ subgroup.  This is the reflection in the formalism of \S2 that
Calabi-Yau sigma models are actually related not to the Landau-Ginzburg
theory of \sup, but to an {\it orbifold} of \sup\ by
\eqn\orb{\ex{-2\pi i \oint J(z)}~.}
So in computing the spectrum of the theory \sup, we will also be
interesting in finding states from various twisted sectors.

In fact, the Landau-Ginzburg orbifold of interest is not quite the
orbifold obtained by projecting onto integral values of \orb .
We are interested in considering both
the (R,R) and the (NS,R) sectors, which yield
spacetime fermions (the (R,NS) and (NS,NS)
sectors, which give the spacetime bosons, then follow by
supersymmetry).  Taking account of the extra free-fermions
$\lambda^{I}$ of \extra, we find that
we are actually interested in projecting onto states with $g=1$ where
\eqn\lorb{g = \exp(-i\pi J_{0})\times (-1)^{\lambda}}
and $\lambda$ denotes the number of $\lambda^{I}$ excitations.
There are then sectors twisted by $g^{k}$ for $k=0,\dots,2m$; for
even $k$, we call them (R,R) sectors (and in particular put the fermions
\extra\ in their Ramond sector) and for odd $k$, we call them (NS,R)
sectors (and give anti-periodic boundary conditions to \extra ).

Similarly, the GSO-projection for right-moving
Ramond sector ends up correlating states with
$\bar q + 3/2$ even with {\it left-handed spacetime} fermions and states with
$\bar q + 3/2$ odd with {\it right-handed spacetime} fermions.
The right-moving $U(1)$ charge also indicates the type of
spacetime supermultiplet a state represents; states
with $\bar q=\pm {1\over 2}$ are parts of chiral or
antichiral superfields in
spacetime, while states with $\bar q=\pm {3\over 2}$ are
parts of vector superfields.

The effect of the GSO-projection in left-moving Ramond sectors is best
summarized separately for models with $E_{6}, ~SO(10)$ and $SU(5)$ gauge
group.  The $E_{6}$ case has been discussed extensively in \Us, so here
we limit ourselves to discussing $SO(10)$ and $SU(5)$ theories.

For $SO(10)$ theories, in Ramond
sectors one must account for the zero modes of the
free fermions \extra\ which realize the $SO(8)$ current algebra.
If we assemble these pairwise into four complex fermions,
their zero modes make the ground state a $16$ dimensional representation
of $SO(8)$.  This is in fact a reducible representation; it splits into
the spinor $\oplus$ spinor$'$ representation of $SO(8)$,
${\bf 16} =  8^{s} \oplus 8^{s'}$.
Then the GSO projection associates
states with $q-2$ {\it odd} with $8^{s}$s of $SO(8)$ and states with
$q-2$ {\it even} with $8^{s'}$s of $SO(8)$.
The $SO(8)\times U(1)$
quantum numbers of the various multiplets which arise in $SO(10)$
theories were summarized in table (1)

For $SU(5)$ theories, the same discussion applies almost verbatim, the
only changes being that the 3 complex fermions provide a reducible $
8$ dimensional representation of $SO(6)$ which splits as $4 \oplus
\bar 4$. Then the GSO projection associates states with
$q-{5\over 2}$ {\it
odd}  with $\bar 4$s of $SO(6)$ and states with $q - {5\over 2}$
{\it even} with $4$s of $SO(6)$.  The $SO(6)\times U(1)$ quantum numbers
of the multiplets which arise in $SU(5)$ theories were also summarized
in table (1).

In order to determine the spectrum of states from twisted sectors, we
must also know the quantum numbers of the ground states.  Because the
fermions have twisted boundary conditions, the ground state of a twisted
sector generically has fractional fermion numbers and non-vanishing
$U(1)$ charges.  Obvious modification of the formulae in \Us\ yields
the following for the $U(1)$ charges of the ground state of the $g^{k}$
twisted sector:
\eqn\groundr{\eqalign { \bar q_{k} =
&\sum_{i} (q_{i}-1) \left( {{k(q_{i}-1)}\over 2} +
\left[ {{k(1-q_{i})}\over 2} \right] + {1\over 2}\right) +
\sum_{a} q_{a} \left( -{k q_{a}\over 2} + \left[ {k q_{a}\over 2}\right] +
{1\over 2}\right) \cr
&+ \sum_{j} q_{j} \left( -{k q_{j}\over 2} + \left[ {k q_{j}\over 2}\right] +
{1\over 2}\right)\cr}}
\eqn\groundl{\eqalign { q_{k} =
&\sum_{i} q_{i} \left( {{k(q_{i}-1)}\over 2} +
\left[ {{k(1-q_{i})}\over 2} \right] + {1\over 2}\right) +
\sum_{a} (q_{a}-1) \left( -{k q_{a}\over 2} + \left[ {k q_{a}\over
2}\right] + {1\over 2}\right) \cr
&+ \sum_{j} (q_{j}-1) \left( -{k q_{j} \over 2} + \left[ {k q_{j}\over
2}\right] + {1\over 2}\right)~. \cr}}
Defining the $\theta_{i,k}$, $\theta_{a,k}$ and $\theta_{j,k}$ to be the
phases present in \groundl\ and \groundr\ (basically,
the twist on the fields relative to
being anti-periodic), these formulae simplify to
\eqn\sgroundr{\bar q_{k} = \sum_{i} (q_{i}-1)\theta_{i,k} + \sum_{a}
q_{a} \theta_{a,k} + \sum_{j} q_{j} \theta_{j,k}}
\eqn\sgroundl{q_{k} = \sum_{i} q_{i}\theta_{i,k}
+ \sum_{a} (q_{a}-1)\theta_{a,k} + \sum_{j} (q_{j}-1)\theta_{j,k}~.}

We also need to determine the ground state eigenvalues of $L_{0}$ ($\bar
L_{0}$ vanishes by right-moving supersymmetry).  These are easily
determined using the normal formulas for the energy of a free twisted
boson or fermion (and remembering to include the extra free $E_{8}$ and
$SO(16-2{\tilde r})$ fermions).
It turns out that the vacuum energy in the $k$th
sector for {\it even} $k$ is given by:
\eqn\renergy{E_R = -{{(\tilde r -3)}\over 8}
 + \sum_{a} {\theta_{a,k}^{2} \over 2}
+ \sum_{j} {\theta_{j,k}^{2} \over 2} - \sum_{i} {\theta_{i,k}^{2}\over
2}~.}
For {\it odd} $k$, the formula becomes:
\eqn\nenergy{E_{NS} = -{5\over 8} + \sum_{a} {\theta_{a,k}^{2} \over 2}
+ \sum_{j} {\theta_{j,k}^{2} \over 2} - \sum_{i} {\theta_{i,k}^{2}
\over 2}~.}
One notes that even for the (R,R) sectors, the vacuum
energy will generically no longer vanish; in order to find even the
spectrum of generations and anti-generations, one must
keep the lowest excited modes of the various fields.

\subsec{The Born-Oppenheimer Approximation and $\bar Q_{+,L}$
cohomology}

As in \Us, because we are looking for massless states in spacetime, we
can employ a Born-Oppenheimer approximation and truncate the quantum
fields to their lowest excited modes.  The right-moving N=2
supersymmetry algebra is
\eqn\rightsup{\{\bar Q_{-},\bar Q_{+}\} = \bar L_{0}, ~~~~\bar Q_{-}^{2}
= \bar Q_{+}^{2} = 0 ~.}
We are interested in finding massless states, which are annihilated by
$\bar L_{0}$; by \rightsup, finding such states
is equivalent to calculating the
cohomology of the $\bar Q_{+}$ operator.  Actually,
we are interested in the subspace of the $\bar Q_{+}$ cohomology which
is also annihilated by $L_{0}$; since in twisted sectors $L_{0}$
generally does not annihilate the vacuum (see \renergy,
\nenergy), we generally need to consider excitations over the vacuum by
the lowest modes of the various fields.\foot {Sectors with {\it
positive} vacuum eigenvalue obviously contribute no massless states, and
do occasionally occur.}

Fortunately, an explicit and
useful expression for the generators of the
N=2 algebra of Landau-Ginzburg models, and
in particular the $\bar Q_{+}$ operator, has recently been provided in
\WitMin\ (see also \Fre, where a closely related construction was
given):
\eqn\barq{\bar Q_{+} = i \oint ~\left( i\bar\psi^{i}\bar\partial
\phi_{i} + {\cal W}\vert_{\theta = 0} \right)
{}~.}
For purposes of computing the cohomology of \barq, it is convenient to
divide it further into
\eqn\splitq{\bar Q_{+,L} = i\oint ~
{\cal W}\vert_{\theta = 0}, ~~~~~\bar Q_{+,R} =
i\oint~i\bar\psi^{i}\bar\partial \phi_{i}~.}

Then, because rescaling the superpotential ${\cal W} \rightarrow \epsilon
{\cal W}$
amounts to a modification of the kinetic energy, which is $\bar Q_{+}$
exact and cannot affect the $\bar Q_{+}$ cohomology, one may compute the
cohomology of \barq\ treating $\bar Q_{+,L}$ as a perturbation.
The
formal justification of this argument is given in \Us.
The result is that the cohomology of $\bar Q_{+}$ is equal to the
cohomology of $\bar Q_{+,L}$ in the cohomology of $\bar Q_{+,R}$.
The cohomology of $\bar Q_{+,R}$ is just the space
of states with holomorphic
dependence on bosonic zero modes and no right-moving excitations;
furthermore, the $\psi^{i}$ and $\bar \psi^{i}$
zero modes can be omitted.
So in evaluating the $\bar Q_{+}$ cohomology of models, we will in
practice evaluate the cohomology of $\bar Q_{+,L}$ in the truncated
Hilbert space of states which form the $\bar Q_{+,R}$ cohomology.

\subsec{Stability, World Sheet Instantons, and \LG\ Theory}

It has long been believed that the generic (0,2) theory based on a stable
bundle $E$ does not exist because of worldsheet instanton effects \DSWW, but
the criterion which distinguished those theories which do exist from those
which don't has proven elusive. Can \LG\ theory help?

First let us consider (0,2) theories which are deformations of (2,2) theories.
In the class of models we have been discussing, these are rank $\tilde r=4$
bundles $E$ on hypersurfaces in $\wp{4}$, where $\{n_i\}=\{w_i\}$. When
\eqn\edefodat{F_i(\phi)=\partial_i W(\phi)}
we have the original (2,2) theory. There
are extra gluinos (states with $\bar q=-3/2$) which fill out the adjoint of
$E_6$. In $SO(8)$ language, these are an $8^{s}_{-1},\ 8^{s}_1$ from the R
sector, and $8^v_1,\ 8^v_{-1},\ 1_0$ from the NS sector. Clearly, we can
deform this theory in a way that breaks (2,2) supersymmetry, but leaves $E_6$
unbroken, by generalizing \edefodat\
 to have $W$ be in the {\it ideal} generated
by the $F_i$s.

Can we deform further and actually break $E_6\to SO(10)$? Clearly, this is
possible only if we can give masses to the extra gluinos through the Higgs
mechanism. In the current language, we want the corresponding states to
cease to be in the $\bar Q_{+,L}$ cohomology by pairing up with states at
$\bar q=-1/2$. These are matter fields. For example, the  $8^{s}_{-1}$
should pair up with an $8^{s}_{-1}$ at $\bar q=-1/2$ which is part of a {\bf
27} of $E_6$, and the $8^{s}_1$ gluino should pair up with an $8^{s}_1$ which
is part of a $\overline{\bf 27}$.

In the case of the $8^{s}_{-1}$, which occurs in the untwisted ($k=0$)
sector, this is never a problem. However, for the $8^{s}_1$, we now see
that there can be an obstruction. For though there may indeed be an $8^{s}_1$
matter field lurking around, it need not occur in the same twisted sector
as the $8^{s}_1$ gluino. It is then impossible for them to pair up. This
would spoil the quantum symmetry of the LG theory. Thus further (0,2)
deformations which break $E_6$ are forbidden.

This is not an idle curiosity. It occurs, for example for the quintic in
$\cp{4}$. There the $8^{s}_1$ gluino occurs in the $k=2$ sector, but the
$8^{s}_1$ matter field occurs in the $k=4$ sector. Thus, contrary to
expectations,
the (0,2) $SO(10)$ deformation of the (2,2) model on the quintic described in
 \twozero\
 does not exist\foot{This is not to say that the $E_{6}$-breaking flat
directions found in \miracles\ do not exist. Rather, if they do exist, the
corresponding (0,2) SCFTs are not describable as the low-energy limit of the
microscopic lagrangians in \S2. It would be very interesting to find the
worldsheet SCFTs which {\it do} describe these deformations.}.

A more {\it recherch\'e} example is provided by the hypersurface
$Y_{W4;10}$ in $\wp{4}_{1,1,1,2,5}$, and a rank 4 bundle defined by
$\{n_a\}=\{1,1,1,1,7\}$. This example will be discussed in more detail in
\S4.2. If one takes $W$ to be of width 1
in the ideal generated by the
$F_a$s \foot{The {\it width} of $W$ is the minimum degree of the polynomial
$Q(\phi)$ such that $QW$ is in the ideal generated by the $F_a$s.},
this is actually
an $E_6$ theory which is {\it not} a deformation of a (2,2) theory. There are
165 {\bf 27}s, whose $8^{s}_{-1}$  components lie in the $k=0$ sector and a
$\overline{\bf 27}$, whose $8^{s}_1$ components is the ground state  of the
$k=6$ sector. This is in agreement \foot{Note that there is a mistake in the
table in \Greene\
listing the number of generations for this model.} with the index, $\half
c_3(E)=164$. Unfortunately, the
$8^{s}_1$ gluino is the state $\bar\phi^5_{-{1\over11}}\ket{0}$ in the $k=4$
sector. It  cannot pair up with the  $\overline{\bf 27}$ without spoiling the
quantum symmetry, so we are {\it not allowed} to deform this theory to generic
values of the data, where it would be an $SO(10)$ theory.

Thus we have a simple criterion for determining whether a would-be $SO(10)$
theory is really an $E_6$ theory. Simply look for extra gluinos coming from
twisted sectors (it suffices to look for  $8^{s}_1$ gluinos coming from $k$
even sectors). If we find one, and there isn't an $8^{s}_1$ matter
field from the same twisted sector for it to pair up with, then we must
choose the defining data in such a way that we in fact get an $E_6$ theory.

Note that though this argument was phrased at the \LG\ point, where the quantum
symmetry is unbroken, it still holds when we move away from the LG point by
turning on a VEV for the twist field (the \Ka\ modulus).
Though the quantum symmetry is
spontaneously broken (by the VEV of the twist field), the Higgs mass term
should
still transform as some definite representation of the quantum symmetry. But
this is impossible, since the part that comes from the $k=0$ sector is a
singlet, whereas the part which pairs up the $8^{s}_1$s transforms nontrivially
under the quantum symmetry.

It would be very interesting to understand  the manifestation of this
phenomenon on the \cy\ side. Presumably, the $E_6$ theories for special
values of the defining data are fine, but when one deforms them to generic
values of the defining data, the resulting $SO(10)$ theories are
destabilized by worldsheet instantons.

The statement, which seems quite remarkable, is that for the special values
of the defining data, the $F_a$s are such that, when restricted to any rational
curve $C$ on $M$, they become linearly dependent (more precisely, one of them
ends up in the ideal generated by the other 4, viewed now as elements of the
polynomial ring $\BC\bigl[\H{0}{\CO_C(1)}\bigr]$). This is precisely the
condition\foot{Equivalently, it is the condition that
$\H{0}{E|_C\otimes\CO_C(-1))}\neq 0$.}
that the bundle $E$ defined by \efundseq\ splits nontrivially when
restricted to $C$. When we deform the $F_a$s, however, the resulting bundle
$E$ has trivial splitting type and the model is destabilized by worldsheet
instantons  \refs{\twozero,\DSWW}.

Perhaps even more remarkable from the worldsheet instanton point of view,
is that there are models in which {\it generic} choices of the $F_a$s lead to
stable theories. In the \LG\ description, however, there is nothing remarkable
about them at all. There are simply no extra gluinos coming from twisted
sectors to mess things up. We will look in detail at one such model in \S4.1.

\newsec{Examples}

In
this section, we present
some examples of (0,2) Landau-Ginzburg vacua in
greater detail.  In \S4.1 we provide a  detailed
discussion of an $SO(10)$ theory
on a complete intersection in
$\wp{5}$.  In somewhat less detail, we discuss in \S4.2 the example in
$\wp{4}_{1,1,1,2,5}$ which was mentioned in \S3.3.

\subsec{The Complete Intersection $Y_{W5;4,4}$}

Consider the (0,2) model  which consists of
a complete intersection
Calabi-Yau manifold in $\wp{5}_{1,1,1,1,2,2}$ defined by two equations
of degree four (so $d_1=d_2=4$), and a rank four vacuum gauge bundle with
$\{n_{a}\} = \{1,1,1,1,1\}$ (so $m=5$). This is the $Y_{W5;4,4}$ (0,2) model
listed in \Greene.

Following the formalism of \S2,3 we see that there should be a
Landau-Ginzburg phase with field content and $U(1)$ charges given
in table (3).

$x$ and $\omega$ are the lowest components of the ``spectator''
superfields; since they appear quadratically in $\bar Q_{+}$, they drop
out in the computation of the massless spectrum and we ignore them
henceforth.

Since $m=5$, this model is a $\BZ_{5}$ orbifold and there are
sectors twisted by
$g^{k}$ for $k=0,\dots 9$.  The vacuum energies and $U(1)$ charges of
sectors $k=0,\dots 5$ are given in table (4).
The vacuum quantum numbers and spectra arising from higher $k$
sectors can be inferred from those of the $k=0,\dots
5$ sectors by CPT.

\overfullrule=0pt
\def\tablerule{\multispan{6}{\tabskip=0pt\hrulefill}&\cr}
\def\tablepad{height3pt&&&&&&\cr}
$$\vbox{\offinterlineskip\tabskip=0pt\halign{
&\vrule#&\quad\hfil $#$ \hfil\quad &\vrule#&\quad\hfil $#$
\hfil \quad&\vrule #& \quad\hfil $#$ \hfil \quad&\vrule#\cr
\omit&&\omit&\hbox{Left $U(1)$
}&\omit&\hbox{Right $U(1)$}&\omit\cr
\omit&\hbox{Field}&\omit&\hbox{charge $q$
}&\omit&\hbox{charge $\bar q$}&\omit\cr
\noalign{\smallskip}
\tablerule\tablepad
& \phi_{1,2} &&{2\over 5}&&{2\over 5}&\cr
\tablepad\tablerule\tablepad
&\phi_{3,\dots 6}&&{1\over 5}&&{1\over 5}&\cr
\tablepad\tablerule\tablepad
&\psi^{1,2}&&{2\over 5}&&-{3\over 5}&\cr
\tablepad\tablerule\tablepad
&\psi^{3,\dots 6}&&{1\over 5}&&-{4\over 5}&\cr
\tablepad\tablerule\tablepad
&\lambda^{1,\dots 5}&&-{4\over 5}&&{1\over 5}&\cr
\tablepad\tablerule\tablepad
&\sigma^{1,2}&&-{4\over 5}&&{1\over 5}&\cr
\tablepad\tablerule\tablepad
&x&&{2\over 5}&&{2\over 5}&\cr
\tablepad\tablerule\tablepad
&\omega&&-{2\over 5}&&{3\over 5}&\cr
\tablepad\tablerule
\noalign{\bigskip}
\noalign{\hsize=2.375in\noindent{\bf Table 3:}
Left and right $U(1)$ charges of fields in
Landau-Ginzburg phase of $Y_{W5;4,4}$.}
 }}
\ifx\answ\bigans\hskip.5in\else\quad\fi
\vbox{\offinterlineskip\tabskip=0pt\halign{
&\vrule#&\quad\hfil $#$ \hfil\quad &\vrule#&\quad\hfil $#$
\hfil \quad&\vrule #& \ \hfil $#$ \hfil \ &\vrule#\cr
\omit&&\omit&\hbox{Vacuum $U(1)$
}&\omit&\hbox{Vacuum}&\omit\cr
\omit&k&\omit&\hbox{Charges ($q,\bar q$)
}&\omit&\hbox{Energy}&\omit\cr
\noalign{\smallskip}
\tablerule\tablepad
&0&&(-2,-{3\over 2})&&0&\cr
\tablepad\tablerule\tablepad
&1&&(0,-{3\over 2})&&-1&\cr
\tablepad\tablerule\tablepad
&2&&(2,-{3\over 2})&&0&\cr
\tablepad\tablerule\tablepad
&3&&(-{8\over 5},-{1\over 10})&&-{2\over 5}&\cr
\tablepad\tablerule\tablepad
&4&&({2\over 5},-{1\over 10})&&-{1\over 5}&\cr
\tablepad\tablerule\tablepad
&5&&(-{12\over 5},{1\over 10})&&0&\cr
\tablepad\tablerule
\noalign{\bigskip}
\noalign{\hsize=2.125in\noindent{\bf Table 4:}
Quantum numbers of twisted sector vacua in Landau-Ginzburg phase of
$Y_{W5;4,4}$.}
 }}$$

The sectors of origin and
$SO(8)\times U(1)$ quantum numbers of the
the left-handed gauginos of $SO(10)$
(assuming a ``generic'' choice of defining data)
are shown in table
(5).   The right-handed gauginos come from sectors $k=0,8,9$ and
are related to the above by CPT. For ``non-generic'' choices of the defining
data one can obtain extra $U(1)$ gauginos from the $k=1$ sector; these are
similar to  the extra $U(1)$ gauge symmetries that arise in Gepner models.
Such gauginos are always accompanied by extra massless
higgsinos, which give them a mass as one moves away from the point of
enhanced symmetry.

\def\tablerule{\multispan{6}{\tabskip=0pt\hrulefill}&\cr}
\def\tablepad{height3pt&&&&&&\cr}
$$\hskip .5in\vbox{\offinterlineskip\tabskip=0pt\halign{
&\vrule#&\hfil $#$ \hfil &\vrule#&\quad\hfil $#$
\hfil \quad&\vrule #&\quad  $#$ \hfil \quad&\vrule#\cr
\omit&\hbox{$SO(8)\times U(1)$ rep.}&\omit&k&\omit&\hbox{State}&\omit\cr
\noalign{\smallskip}
\tablerule\tablepad
&8^{s'}_{-2}&&0&&\ket{0}&\cr
\tablepad\tablerule\tablepad
&28_0&&1&&\lambda^I_{-{1\over2}}\lambda^J_{-{1\over2}}\ket{0}&\cr
\tablepad\tablerule\tablepad
&1_0&&1&&
\Bigl(\sum_iw_i\phi^i_{-{q_i\over2}}\bar\phi^i_{-1+{q_i\over2}}
+4\sum_a\lambda^a_{-{3\over5}}\bar\lambda^a_{-{2\over5}}&\cr  \tablepad
&&&&&\hfill+4\sum_j\sigma^j_{-{3\over5}}\bar\sigma^j_{-{2\over5}}\Bigr)\ket{0}
&\cr
\tablepad\tablerule\tablepad
&8^{s'}_{2}&&2&&\ket{0}&\cr
\tablepad\tablerule
\noalign{\bigskip}
\noalign{\hsize=4in\noindent{\bf Table 5:}
The left-handed {\bf 45} ($\bar q=-3/2$) in the \LG\ phase of $Y_{W5;4,4}$.}
 }}
$$
\bigskip

The generations, which are {\bf 16}s of $SO(10)$, arise as shown
in table (6).  There are generically {\it no} anti-generations.
The degree 5 polynomial $P(\phi)$ is subject to the equivalence relation
\eqn\equivrelPfive{P_5(\phi)\sim P_5(\phi)+c^a F_a(\phi)+
s^j(\phi)W_j(\phi)\quad.}
There are 80 independent degree five polynomials modulo these relations, in
agreement with the expected number of generations $\half c_3(E)$.

\def\tablerule{\multispan{6}{\tabskip=0pt\hrulefill}&\cr}
\def\tablepad{height3pt&&&&&&\cr}
$$\hskip.5in\vbox{\offinterlineskip\tabskip=0pt\halign{
\vrule#&\hfil $#$ \hfil &\vrule#&\quad\hfil $#$
\hfil \quad&\vrule #&\quad  $#$ \hfil \quad&\vrule#\cr
\omit&\hbox{$SO(8)\times U(1)$ rep.}&\omit&k&\omit&\hbox{State}&\omit\cr
\noalign{\smallskip}
\tablerule\tablepad
&8^{s}_{-1}&&0&&P_5(\phi^i_{\scriptscriptstyle 0})\ket{0}&\cr
\tablepad\tablerule\tablepad
&8^v_1&&1&&\lambda^I_{-{1\over2}}P_5(\phi^i_{-{q_i\over2}})\ket{0}&\cr
\tablepad\tablerule
\noalign{\bigskip}
\noalign{\hsize=3in\noindent{\bf Table 6:}
Right-handed {\bf 16}s ($\bar q=-1/2$) in the \LG\ phase of $Y_{W5;4,4}$.}
 }}
$$

\bigskip
The left-handed {\bf 16}$'$s which are the antiparticles of these arise in a
slightly unobvious fashion. The $8^s_1$ component arises as 80 states in the
$k=0$ sector of the form $P_{11}(\phi)\bar\lambda^a\ket{0},
P_{11}(\phi)\bar\sigma^j\ket{0}$ which are in the cohomology of $\bar Q_{+,L}$.
The $8^v_{-1}$ component arises as 80 states in the $k=9$ sector of the form
$\lambda^IP_5(\bar\phi)\ket{0}$.

The left- and right-handed {\bf 10}s of
$SO(10)$ arise as shown in table (7).
The degree 10 polynomial $P_{10}(\phi)$ is subject to the equivalence
relation
\eqn\epten{P_{10}(\phi) \sim P_{10}(\phi) + c^{a}(\phi)F_{a}(\phi)
+ s^{j}(\phi)W_{j}(\phi)~.}
Generically, there are 72 independent degree 10 polynomials $P_{10}(\phi)$
modulo these relations. Two more
{\bf 10}s arise from  twisted sectors. Slightly subtle are the $1_2$
components of the left-handed {\bf 10}s, which arise from the $k=5$ sector.
Though this is a NS sector, the left-moving fermions are untwisted.
There are 42 states of the form $\phi^{1,2}_{\scriptscriptstyle0}\times$a
fifth order polynomial in $\{\bar\lambda^1_{\scriptscriptstyle0},
\dots,\bar\lambda^5_{\scriptscriptstyle0},\bar\sigma^1_{\scriptscriptstyle0},
\bar\sigma^2_{\scriptscriptstyle0}\}$. Of these, generically, two are
in the cohomology of $\bar Q_{+,L}$.
All in all, we find 74 left-handed and 74 right-handed ${\bf 10}$s of $SO(10)$.

\def\tablerule{\multispan{10}{\tabskip=0pt\hrulefill}&\cr}
\def\staterule#1{
&#1&&\omit\hrulefill&&\omit\hrulefill&&\omit\hrulefill&&\omit\hrulefill&\cr}
\def\tablepad{height3pt&&&&&&&&&&\cr}
$$\vbox{\offinterlineskip\tabskip=0pt\halign{
\vrule#&\hfil $#$ \hfil&\vrule#&
\quad\hfil $#$ \hfil\quad &\vrule#&\quad\hfil $#$ \hfil\quad
&\vrule#&\quad\hfil $#$ \hfil \quad&\vrule #&\quad  $#$ \hfil \quad&\vrule#\cr
\omit&&\omit&\multispan{3}\hbox{\ Left-handed
($\bar q={1\over2}$)}&\omit
&\multispan{3}\hbox{\ Right-handed ($\bar q=-{1\over2}$)}&\omit\cr
\noalign{\smallskip} \omit&&\multispan{8}\hrulefill&\cr
\omit&&height3pt&&\omit&&&&\omit&&\cr
\omit&\hbox{$SO(8)\times U(1)$
rep.}&&k&\omit&\hbox{State}&&k&\omit&\hbox{State}&\cr
\omit&&height3pt&&\omit&&&&\omit&&\cr
\tablerule\tablepad
&&&0&&P_{10}(\phi^{i}_{\scriptscriptstyle
0})\ket{0}&&0&&
P_6(\phi^i_{\scriptscriptstyle0})\bar\lambda^a_{\scriptscriptstyle0}\ket{0},
&\cr
\tablepad
&8^{s'}_0&&&&&&&&
P_6(\phi^i_{\scriptscriptstyle0})\bar\sigma^j_{\scriptscriptstyle0}\ket{0}
&\cr
\tablepad\staterule{}\tablepad
&&&6&&\phi^{1,2}_{-{1\over 5}}\ket{0}&&
4&&\bar\phi^{1,2}_{-{1\over 5}}\ket{0}&\cr
\tablepad\tablerule\tablepad
&&&1&&P_{10}(\phi^{i}_{-{q_{i}\over2}})\ket{0}&
&1&&P_{6}(\phi^{i}_{-{q_{i}\over2}})\bar\lambda^a_{-{2\over5}}\ket{0},&\cr
&1_{2}&&&&&&&&P_{6}(\phi^{i}_{-{q_{i}\over2}})\bar\sigma^j_{-{2\over5}}\ket{0}
&\cr
\tablepad\staterule{}\tablepad
&&&7&&\phi^{1,2}_{-{2\over 5}}\ket{0}&
&5&&\phi^{1,2}_{\scriptscriptstyle0}
\prod_5\{\bar\lambda^a_{\scriptscriptstyle0},
\bar\sigma^j_{\scriptscriptstyle0}\}\ket{0}&\cr
\tablepad\tablerule\tablepad
&&&9&&P_{6}(\bar\phi^{i}_{-{q_{i}\over2}})\lambda^a_{-{2\over5}}\ket{0},&
&9&&P_{10}(\bar\phi^{i}_{-{q_{i}\over2}})\ket{0}&\cr
&1_{-2}&&&&P_{6}(\bar\phi^{i}_{-{q_{i}\over2}})\sigma^j_{-{2\over5}}\ket{0}
&&&&&\cr
\tablepad\staterule{}\tablepad
&&&5&&\phi^{1,2}_{\scriptscriptstyle0}\ket{0}&
&3&&\bar\phi^{1,2}_{-{2\over5}}\ket{0}&\cr
\tablepad\tablerule
\noalign{\bigskip}
\noalign{\hsize=4in\noindent{\bf Table 7:}
${\bf 10}$s  in the \LG\ phase of $Y_{W5;4,4}$.}
 }}
$$

\bigskip

We can also find the spectrum of gauge singlet
matter fields, which always originate
in the (NS,R) sectors.
The right-handed singlets are shown in table (8).
The states in the $k=1$ sector should be considered
modulo the equivalence relations
\eqn\epfourequi{
\eqalign{&F_{a}(\phi^{i}_{-{q_{i}\over 2}})
\lambda^{b}_{-{3\over 5}}\ket{0}  \sim F_{a}(\phi^{i}_{-{q_{i}\over 2}})
\sigma^{j}_{-{3\over 5}}
\ket{0} \sim 0 \cr
&W_{j}(\phi^{i}_{-{q_{i}\over 2}})
\lambda^{a}_{-{3\over 5}}\ket{0}
\sim W_{j}(\phi^{i}_{-{q_{i}\over 2}})\sigma^{k}_{-{3\over 5}}\ket
{0} \sim 0 \cr
&P_{2}(\phi^{i}_{-{q_{i}\over 2}})
\left[{\partial F_{a}\over \partial \phi^{1,2}_{-{1\over 5}}}
\lambda^{a}_{-{3\over 5}}
+ {\partial W_{j}\over \partial \phi^{1,2}_{-{1\over 5}}}
\sigma^{j}_{-{3\over 5}} \right] \ket{0}
\sim 0 \cr
&P_{1}(\phi^{3,\dots 6}_{-{1\over 10}})
\left[{\partial F_{a}\over \partial \phi^{3,\dots
6}_{-{1\over 10}}}\lambda^{a}_{-{3\over 5}} +
{\partial W_{j}\over \partial \phi^{3,\dots
6}_{-{1\over 10}}}\sigma^{j}_{-{3\over 5}}\right] \ket{0} \sim 0~.}}
Subjecting the states in the $k=1$ sector to the equivalence relations
\epfourequi, we find 318 gauge singlets.  Adding to them the 21 obtained
from quadratic, antisymmetric combinations of the $\lambda^{a}_{-{1\over
5}}$ and
$\sigma^{j}_{-{1\over 5}}$ oscillators in
the $k=3$ sector, we find a total of 339 right-handed gauge singlets in
this model.

\def\tablerule{\multispan{6}{\tabskip=0pt\hrulefill}&\cr}
\def\tablepad{height3pt&&&&&&\cr}
$$\hskip.25in\vbox{\offinterlineskip\tabskip=0pt\halign{
\vrule#&\quad\hfil $#$ \hfil\quad &\vrule#&\quad\hfil $#$
\hfil \quad&\vrule #&\quad  $#$ \hfil \quad&\vrule#\cr
\omit&\hbox{$SO(8)\times U(1)$ rep.}&\omit&k&\omit&\hbox{State}&\omit\cr
\noalign{\smallskip}
\tablerule\tablepad
&1_{0}&&1&&P_4(\phi^i_{-{q_{i}\over 2}})\lambda^{a}_{-{3\over5}}\ket{0},~
P_{4}(\phi^{i}_{-{q_{i}\over 2}})\sigma^{j}_{-{3\over5}}\ket{0}&\cr
\tablepad\tablerule\tablepad
&1_{0}&&3&&\lambda^{a}_{-{1\over 5}}\lambda^{b}_{-{1\over
5}}\ket{0},\  \lambda^{a}_{-{1\over 5}}\sigma^{j}_{-{1\over 5}}\ket{0},\
\sigma^{1}_{-{1\over 5}}\sigma^{2}_{-{1\over 5}}\ket{0}&\cr
\tablepad\tablerule
\noalign{\bigskip}
\noalign{\hsize=4in\noindent{\bf Table 8:}
Right-handed singlets ($\bar q=-1/2$) in the \LG\ phase of $Y_{W5;4,4}$.}
 }}
$$

\bigskip

\leftline{\it Comparison with \cy\ expectations}

How does the spectrum we have found compare with that expected from the
\cy\ side? The {\bf 16}s agree both in number (80), and in
deformation-theoretic representation \twozero. From the definition of $E$
\efundseq, it is clear that $\H{1}{E}$ has a deformation-theoretic
representation as degree 5 polynomials {\it modulo} $\{F_a(\phi),W_j(\phi)\}$.

Of course, the {\it net} number of generations ({\bf 16}s minus {\bf 16}$'$s)
is protected by an index theorem, and must be the same in the two phases. Not
so the number of {\bf 10}s. A mass term for {\bf 10}s is not forbidden in the
spacetime
superpotential, and we might expect their number to jump as we move about in
the moduli space.

 Taking the second exterior power of \efundseq, one finds
that $\H{2}{\bigwedge^2E}$ (which is Serre-dual to $\H{1}{\bigwedge^2E}$) has
a deformation-theoretic representation as degree 10 polynomials {\it modulo}
$\{F_a(\phi),W_j(\phi)\}$\foot{One also finds that $\H{1}{\bigwedge^2E}$ is
isomorphic to the cohomology of the sequence \hfil\break
$\H{0}{\CO(2)^{\oplus10}}{\buildrel u\over\longrightarrow}
\H{0}{\CO(6)^{\oplus5}}{\buildrel v\over\longrightarrow}\H{0}{\CO(10)}$, where
$u(s_{abc})=\epsilon^{abcde}s_{abc}F_d$ and $v(t^a)=t^a F_a$. After some
work, one finds that this gives precisely the
expressions that we found for 72
of the right-moving {\bf 10}s
in terms of degree six polynomials in $\phi$.}.
 This accounts for 72
of the 74 {\bf 10}s we have found. The other two (from the $k=5,6,7$ sectors)
have no analogues in the \cy\ analysis. It would be interesting to see  them
pick up a mass as we move away from the \LG\ point.

Finally, we turn to the singlets.
In the Landau-Ginzburg
phase of (2,2) theories, one
is able to further subdivide the gauge singlets
into
deformations of complex structure
(elements of $\H{1}{T}$), deformations of \Ka\ structure (elements of
$\H{1}{T^{*}}$) and deformations of the bundle $E$ (elements of
$\H{1}{End\ E}$, where
$E$ is taken to be the tangent bundle
for a (2,2) model) \Us. Some of the states we have found clearly cry out for
such an interpretation. For instance, the states
$P_{4}(\phi^{i})\sigma^{j}\ket{0}$ in the $k=1$ sector are naturally associated
with deformations of the corresponding polynomial $W_j(\phi)$. However, not
all of the allowed deformations of $W_j$ are accounted for. In particular, we
have modded out by deformations proportional to the $F_a(\phi)$. But the
states  $\lambda^{a}\sigma^{j}\ket{0}$ from the $k=3$ sector precisely
compensate for this error.

Similarly, the \cy\ analysis (using the methods of \Miron)
tells us that an element of $\H{1}{End\ E}$ is
specified by giving five degree 4 polynomials, modulo $\{F_a,W_j\}$. Since
there are 51 independent such polynomials, we {\it expect} to find 255
states corresponding to elements of  $\H{1}{End\ E}$. Natural candidates are
the states $P_{4}(\phi^{i})\lambda^a\ket{0}$ in the $k=1$ sector.

Since $\chi(M)=-144$, the \cy\ analysis has accounted for 73 complex
structure moduli, one \Ka\ modulus, and 255 elements of $\H{1}{End\ E}$, or
329 of the 339 singlets that we have found.

There was no guarantee that we
would come even close to agreement, but we have done surprisingly well. There
are ten extra singlet states in the $k=3$ sector over what was expected from
field theory, and there were two extra {\bf 10}s.

Of course, it is probably wise not to take too seriously the distinction
between moduli which deform the complex structure of $M$, and those
which
deform the holomorphic structure of $E$. In (0,2) \LG\ theory, there is really
no invariant distinction between them. As we shall see in \S5, this opens up
the possibility of topology-changing processes in (0,2) theories which are
impossible in (2,2) theories.

\vfill

\subsec{The Hypersurface $Y_{W4;10}$}

Now we discuss, in somewhat less detail than the previous example, the
(0,2) model which consists of a Calabi-Yau hypersurface in
$\wp{4}_{1,1,1,2,5}$ with a gauge bundle defined by
$\{n_{a}\} =\{1,1,1,1,7\}$.  This model serves as an illustration of the
phenomenon described in \S3.3; it only seems to make sense for highly
restricted choices of the defining data, and is then an $E_{6}$ theory
instead of an $SO(10)$ theory (which is what one would expect generically,
since it has a rank four vacuum gauge bundle).

The Landau-Ginzburg phase has field content and $U(1)$ charges given in
table (9)
(we have omitted the ``spectators'' $x$ and $\omega$ of \S2; they do not
enter in any of the computations).
Since $m=11$, this model is a $\BZ_{11}$ orbifold and there
are sectors twisted by $g^{k}$ for $k=0,\dots 21$.  The vacuum energies
and $U(1)$ charges of sectors $k=0,\dots 11$ are given in table (10).
The content of the other sectors can be inferred by CPT.

\overfullrule=0pt
\def\tablerule{\multispan{6}{\tabskip=0pt\hrulefill}&\cr}
\def\tablepad{height3pt&&&&&&\cr}
$$\vbox{\offinterlineskip\tabskip=0pt\halign{
&\vrule#&\quad\hfil $#$ \hfil\quad &\vrule#&\quad\hfil $#$
\hfil \quad&\vrule #& \quad\hfil $#$ \hfil \quad&\vrule#\cr
\omit&&\omit&\hbox{Left $U(1)$
}&\omit&\hbox{Right $U(1)$}&\omit\cr
\omit&\hbox{Field}&\omit&\hbox{charge $q$
}&\omit&\hbox{charge $\bar q$}&\omit\cr
\noalign{\smallskip}
\tablerule\tablepad
& \phi_{1,2,3} &&{1\over 11}&&{1\over 11}&\cr
\tablepad\tablerule\tablepad
&\phi_{4}&&{2\over 11}&&{2\over 11}&\cr
\tablepad\tablerule\tablepad
&\phi_{5}&&{5\over 11}&&{5\over 11}&\cr
\tablepad\tablerule\tablepad
&\psi^{1,2,3}&&-{10\over 11}&&{1\over 11}&\cr
\tablepad\tablerule\tablepad
&\psi^{4}&&-{9\over 11}&&{2\over 11}&\cr
\tablepad\tablerule\tablepad
&\psi^{5}&&-{6\over 11}&&{5\over 11}&\cr
\tablepad\tablerule\tablepad
&\lambda^{1,\dots,4}&&-{10 \over 11}&&{1\over 11}&\cr
\tablepad\tablerule\tablepad
&\lambda^{5}&&-{4\over 11}&&{7\over 11}&\cr
\tablepad\tablerule\tablepad
&\sigma&&-{10\over 11}&&{1\over 11}&\cr
\tablepad\tablerule
\noalign{\bigskip}
\noalign{\hsize=2.375in\noindent{\bf Table 9:}
Left and right $U(1)$ charges of fields in
Landau-Ginzburg phase of $Y_{W4;10}$.}
 }}
\ifx\answ\bigans\hskip.5in\else\quad\fi
\vbox{\offinterlineskip\tabskip=0pt\halign{
&\vrule#&\quad\hfil $#$ \hfil\quad &\vrule#&\quad\hfil $#$
\hfil \quad&\vrule #& \ \hfil $#$ \hfil \ &\vrule#\cr
\omit&&\omit&\hbox{Vacuum $U(1)$
}&\omit&\hbox{Vacuum}&\omit\cr
\omit&k&\omit&\hbox{Charges ($q,\bar q$)
}&\omit&\hbox{Energy}&\omit\cr
\noalign{\smallskip}
\tablerule\tablepad
&0&&(-2,-{3\over 2})&&0&\cr
\tablepad\tablerule\tablepad
&1&&(0,-{3\over 2})&&-1&\cr
\tablepad\tablerule\tablepad
&2&&(2,-{3\over 2})&&0&\cr
\tablepad\tablerule\tablepad
&3&&(-{6\over 11},-{23\over 22})&&-{9\over 11}&\cr
\tablepad\tablerule\tablepad
&4&&({16\over 11},-{23\over 22})&&-{1\over 11}&\cr
\tablepad\tablerule\tablepad
&5&&(-{7\over 11},-{25\over 22})&&-{13\over 22}&\cr
\tablepad\tablerule\tablepad
&6&&(1,-{1\over 2})&&0&\cr
\tablepad\tablerule\tablepad
&7&&(-{17\over 11},-{1\over 22})&&-{9\over 22}&\cr
\tablepad\tablerule\tablepad
&8&&({5\over 11},-{1\over 22})&&-{2\over 11}&\cr
\tablepad\tablerule\tablepad
&9&&(-{18\over 11},-{3\over 22})&&-{4\over 11}&\cr
\tablepad\tablerule\tablepad
&10&&({4\over 11},-{3\over 22})&&-{2\over 11}&\cr
\tablepad\tablerule\tablepad
&11&&(-{26\over 11},{3\over 22})&&0&\cr
\tablepad\tablerule
\noalign{\bigskip}
\noalign{\hsize=2.125in\noindent{\bf Table 10:}
Quantum numbers of twisted sector vacua in Landau-Ginzburg phase of
$Y_{W4;10}$.}
 }}$$
\bigskip
The Landau-Ginzburg superpotential is given explicitly by
\eqn\eyfoursup{{\cal W} = \Lambda^{a}F_{a}(\phi) + \Sigma W(\phi)}
and as explained in \S3.3, this model {\it does not} define a sensible string
theory for ``generic'' choices of the defining data . (If it did, it {\it
would} be  an $SO(10)$ theory.)  However, it {\it does} make sense
and yields an $E_{6}$ theory on a
subspace of the set of choices for $W$ and the $F_{a}$ where $W$ is
taken to be of width one in the ideal generated by the $F_{a}$. Equivalently,
we can choose $F_1$ to be of width one in the ideal generated by
$W,F_2,\dots,F_5$.  A simple  choice of data defining such a sensible
model is given by
\eqn\eexplicit{\eqalign{
&F_1(\phi)=\phi_{1}\phi_2^{9},\quad F_2(\phi)=\phi_2^{10},\cr
&F_3(\phi)=\phi_3^{10},      \quad F_4(\phi) =\phi_{5}^{2},\cr
&F_{5}(\phi) = \phi_{4}^{2} \cr
}}
with $W(\phi)$ a generic degree 10 polynomial chosen such that
$\{W(\phi)=0\}$ is smooth and does not contain the point $\phi_1\neq0,\
\phi_2=\dots=\phi_5=0$.

The adjoint of $E_6$ decomposes under $SO(10)$ as ${\bf 78}={\bf
16}'\oplus{\bf 45}\oplus{\bf 1}\oplus{\bf 16}$. The states of this model
are, however, more naturally arranged by their quantum numbers under the
$SO(8)\times U(1)$ subgroup of $SO(10)$ which is linearly realized. In table
(11) we list
the left-handed gauginos of $E_{6}$ which come from the R ($k$ even)
sectors. States from NS sectors fill out the rest of the {\bf 78} of $E_{6}$.
These are the $28_0\oplus1_0\oplus1_0\oplus8^v_1$ from the $k=1$ sector and the
$8^v_{-1}$ from the $k=3$ sector.

\def\tablerule{\multispan{6}{\tabskip=0pt\hrulefill}&\cr}
\def\tablepad{height3pt&&&&&&\cr}
$$\hskip .5in\vbox{\offinterlineskip\tabskip=0pt\halign{
&\vrule#&\hfil $#$ \hfil &\vrule#&\quad\hfil $#$
\hfil \quad&\vrule #&\quad  $#$ \hfil \quad&\vrule#\cr
\omit&\hbox{$SO(8)\times U(1)$ rep.}&\omit&k&\omit&\hbox{State}&\omit\cr
\noalign{\smallskip}
\tablerule\tablepad
&8^{s'}_{-2}&&0&&\ket{0}&\cr
\tablepad\tablerule\tablepad
&8^{s}_{-1}&&0&&(\phi^{1}_{\scriptscriptstyle0}
\bar \lambda^2_{\scriptscriptstyle0}
 -
\phi^{2}_{\scriptscriptstyle0}\bar\lambda^1_{\scriptscriptstyle0})\ket{0}&\cr
\tablepad\tablerule\tablepad &8^{s'}_{2}&&2&&\ket{0}&\cr
\tablepad\tablerule\tablepad
&8^{s}_{1}&&4&&\bar\phi^{5}_{-{1\over 11}}\ket{0}&\cr
\tablepad\tablerule
\noalign{\bigskip}
\noalign{\hsize=3in\noindent{\bf Table 11:}
Ramond sector states which form the left-handed {\bf 78} ($\bar q=-3/2$) in
the \LG\ phase of $Y_{W4;10}$.}
 }}
$$
\bigskip

The generations of $E_6$ decompose under $SO(10)$ as ${\bf 27}={\bf
1}\oplus{\bf 16}\oplus{\bf 10}$, which can be further decomposed under
$SO(8)\times U(1)$. In table (12), we display the components of the
right-handed
 $27$s which come from the $k$ even sectors. The polynomials $P_{11}(\phi)$
and $P_{22}(\phi)$ are always defined {\it modulo}
the ideal generated by the $F_{a}(\phi)$ and $W(\phi)$.  For $F_1$ of width 1
in the ideal generated by $F_2,\dots,F_5,W$, there are exactly 165 states of
each type when mod out by the ideal.  So there are 165
{\bf 27}s of $E_{6}$.

There is, in this model, also one right-handed anti-generation, a
$\overline{\bf 27}$ whose components from $k$ even sectors are displayed in
table (13).

\def\tablerule{\multispan{6}{\tabskip=0pt\hrulefill}&\cr}
\def\tablepad{height3pt&&&&&&\cr}
$$\vbox{\offinterlineskip\tabskip=0pt\halign{
&\vrule#&\hfil $#$ \hfil &\vrule#&\quad\hfil $#$
\hfil \quad&\vrule #&\quad  $#$ \hfil \quad&\vrule#\cr
\omit&\hbox{$SO(8)\times U(1)$ rep.}&\omit&k&\omit&\hbox{State}&\omit\cr
\noalign{\smallskip}
\tablerule\tablepad
&8^{s}_{-1}( \subset 16)&&0&&P_{11}(\phi^{i}_{\scriptscriptstyle0})\ket{0}&\cr
\tablepad\tablerule\tablepad
&8^{s'}_{0}(\subset 10)&&0&&P_{22}(\phi^{i}_{\scriptscriptstyle0})\ket{0}&\cr
\tablepad\tablerule
\noalign{\bigskip}
\noalign{\hsize=2.25in\noindent{\bf Table 12:}
Right-handed generations in the \LG\ phase of
$Y_{W4;10}$.}
 }}\quad
\vbox{\offinterlineskip\tabskip=0pt\halign{
&\vrule#&\hfil $#$ \hfil &\vrule#&\quad\hfil $#$
\hfil \quad&\vrule #&\quad  $#$ \hfil \quad&\vrule#\cr
\omit&\hbox{$SO(8)\times U(1)$ rep.}&\omit&k&\omit&\hbox{State}&\omit\cr
\noalign{\smallskip}
\tablerule\tablepad
&8^{s}_{1}( \subset 16')&&6&&\ket{0}&\cr
\tablepad\tablerule\tablepad
&8^{s'}_{0}(\subset 10)&&8&&\bar\phi^{5}_{-{2\over 11}}\ket{0}&\cr
\tablepad\tablerule
\noalign{\bigskip}
\noalign{\hsize=2.25in\noindent{\bf Table 13:}
Right-handed anti-generation in the \LG\ phase of
$Y_{W4;10}$.}
 }}
$$
\bigskip

In particular, the Landau-Ginzburg
theory produces a net of $165-1=164$
generations, in agreement with ${1\over 2}c_{3}(E)$ in the
Calabi-Yau phase.

For ``generic'' choices of the data \eexplicit, the extra
$8^{s}_{-1}$ gaugino from the $k=0$ sector and the extra
$8^{v}_{1}$ and $1_{0}$ gauginos from the $k=1$ sector would be able to
pair up with a matter field (the corresponding components of a {\bf 27}) at
$\bar q=-1/2$ and gain masses via the Higgs mechanism; this is no surprise, the
$k=0,1$ sectors represent exactly the part of the Landau-Ginzburg theory
corresponding to the field theory limit.  But even for generic choices of the
data, it is {\it impossible} for the extra gauginos from the $k=3,4$ sectors to
become massive, because there are no available $\overline{\bf 27}$s in the same
twisted sector.  Thus, though we appear to have a sensible $E_6$ theory with
this choice of the defining data, we cannot deform it to an $SO(10)$ theory.

\newsec{Discussion}

We have seen in \S4
that the (0,2) \LG\ theories have perfectly reasonable spectra, central
charges,
{\it etc.} This is strong evidence that the \LG\ phase indeed
describes fully-fledged (0,2) SCFTs. Since the \cy\ phase is continuously
connected to the LG phase, our conviction that the CY phase is also
well-defined
is strengthened. We have also seen that certain \LG\ theories can be embedded
sensibly in a string theory  only for special values of their defining data
because there is a quantum obstruction to breaking $E_6\to SO(10)$. It remains
to be seen how this restriction is recovered from the consideration of
worldsheet instantons in the \cy\ phase.

The microscopic Lagrangians of \S2 are a powerful tool for exploring the
phase structure of both (2,2) and (0,2) theories. In many cases, knowledge of
the microscopic Lagrangian is enough to determine the global structure of the
\Ka\ moduli space.

The simplest case is the one parameter \Ka\ moduli space with two phases, a
\cy\ phase for $r\gg0$ and a \LG\ phase for $r\ll0$. For a (2,2) model
defined  on a degree $d$  hypersurface in (weighted) $\BP^4$, the
global structure of the moduli space is that of the complex
$a$ plane, minus the points $a^d=1$, modded out by $\BZ_d$ generated by $a\to
\ex{2\pi i/ d}a$. The point $a=0$ corresponds to $r=-\infty$, the
\LG\ point (which has a $\BZ_d$ quantum symmetry \VafaQ). The
points $a^d=1$ all get
mapped into the phase transition point $r=\theta=0$,
and $a=\infty$ into $r=\infty$.
This picture can be verified explicitly using mirror symmetry
\refs{\CandMir,\HWPmirrors}.

For (0,2) theories of the sort we have been discussing, the only difference
in the structure of the \Ka\ moduli space, as predicted by the microscopic
Lagrangian, is that
$d$ is replaced by $m$. For hypersurfaces in $\wp{4}$, we always have those
(0,2) theories which are deformations of the corresponding (2,2) theory, and
hence have $m=d$. But we also have theories, such as the example discussed in
\S4.2, for which $m\neq d$. In these models, the global structure of the  (0,2)
\Ka\ moduli space is {\it different} from that of the corresponding (2,2)
\Ka\ moduli space; in the example of \S4.2, it is a $\BZ_{11}$, rather than a
$\BZ_{10}$ orbifold.

If one generalizes only slightly to complete intersections in
weighted projective space,
the differences become even more pronounced. As we have already seen,
the microscopic theory in
the (2,2) case is in a hybrid phase for $r\ll0$, whereas the (0,2) theory is in
an ordinary \LG\ phase.

Since we don't have a terribly good handle on these
hybrid theories, the microscopic Lagrangian is not much use in determining
the structure
 of the \Ka\ moduli space for (2,2) complete intersections. Mirror symmetry,
however, has yielded results, both for the $Y_{W5;4,4}$ complete intersection
that we discussed in \S4.1, and for the closely related manifold
$Y_{W5;6,6}$, which is the complete intersection of two degree 6 polynomials
in $\wp{5}_{3,3,2,2,1,1}$. Klemm and Theisen, in studying mirror symmetry for
these two manifolds, found the global structure of the corresponding (2,2)
\Ka\ moduli spaces \KT. The
result is that the \Ka\ moduli space again has the
general form
\eqn\estrucka{{\BC - \{a^{d_1+d_2}=1\}\over \BZ_{d_1+d_2}}}
where $d_1+d_2=8$ for $Y_{W5;4,4}$ and $12$ for $Y_{W5;6,6}$.

We have studied
the (0,2) models on both of these manifolds. Details of the former were given
in \S4.1, the latter has $\{n_a\}=\{1,1,1,1,4\}$ \Greene. The \LG\ points
have, respectively, $\BZ_5$ and $\BZ_8$ quantum symmetries. Thus, the
structure
of the \Ka\ moduli space of these (0,2) theories is as in \estrucka, but with
$d_1+d_2$ replaced by $m$, where $m=5,8$ respectively.

At large radius, the identifications in question simply correspond to shifts of
the $B_{\mu\nu}$ field by $2\pi$. What {\it differs} between the (2,2) theories
and (0,2) theories\foot{And, indeed, between different (2,2)
theories as well.} is how this obvious symmetry continues down to small
(negative infinite) radius.

The algebraic geometers, who through the study of mirror symmetry have
only recently become accustomed to thinking about this ``enlarged'' \Ka\
moduli space\foot{Enlarged, both because it contains the moduli of the
$B_{\mu\nu}$ field and because it describes, in addition to the \cy\ phase,
other phases which have no geometrical interpretation. The phrase ``enlarged
\Ka\ moduli space'' was coined
in \refs{\flops,\flopsII}.} are in for a shock.
There is not {\it one} \Ka\ moduli space, but several. There is the (2,2)
moduli space, with which they have become familiar, but
there are also the \Ka\
moduli spaces which arise from (0,2) models and which, in general, have
quite
different global structure.

Perhaps the most exciting feature of the enlarged \Ka\ moduli space of (2,2)
theories to emerge recently is that it may contain {\it several} CY phases
corresponding to {\it topologically distinct} \cy\ manifolds
\refs{\phases,\flops}.
These different manifolds correspond to different ways of resolving the
singularities of a single {\it singular} \cy\ space.

Though we have, in this
paper, confined ourselves to (0,2) models on smooth \cy\ manifolds, nothing
restricted us to this case. The \LG\ theories make perfect sense, even when
the corresponding \cy\ manifold $M$ is singular. In order to compare with the
\cy\ side, however, we need to specify both how the singularities of  $M$
are resolved {\it and} how the sheaf $E$ is lifted to a bundle on
the resolved space.
If there are several different resolutions of $M$, this leads to the phenomenon
of  \refs{\flops,\flopsII}. There may also, however, be distinct liftings
of $E$ to the resolved space \ref\inprogress{Work in progress.},
which leads to the possibility of topologically
distinct (0,2) models leading to the same \LG\ theory.

An even wilder manifestation of topology-change is possible in (0,2) theories.
Recall that at the \LG\ point, there is no invariant distinction between the
fermionic superfields $\Sigma^j$ which multiply the $W_j$ defining the
manifold $M$ and the fermionic superfields $\Lambda^a$ which multiply the
$F_a$ defining the bundle $E$. Therefore, we can contemplate an
automorphism of the (0,2) \LG\ theory which exchanges the roles of the
$\Sigma^j$s and the $\Lambda^a$, and hence of the $W_j$s and the $F_a$s.
Thus {\it topologically distinct} (0,2) models, defined on different manifolds,
may be described by the same \LG\ theory.

An example of this is provided by
complete intersections in $\wp{5}_{1,2,2,2,3,3}$.
Consider the following two theories
\item{1)} $M_1$ is the complete intersection of a degree $d_1=5$ and a
$d_2=8$ hypersurface. $E_1$ is the rank 3 bundle on $M_1$ defined by
$\{n_a\}=\{1,1,2,6\},\ m=10$. So the polynomials
$\{W_1(\phi),W_2(\phi),F_1(\phi),F_2(\phi),F_3(\phi),F_4(\phi)\}$
which define $(M_1,E_1)$ have degrees $\{5,8,9,9,8,4\}$. The
\LG\ superpotential for this theory is
$$\CW=\Sigma^jW_j(\Phi)+\Lambda^aF_a(\Phi)$$
 \item{2)} $M_2$ is the complete intersection of a degree $d_1=4$ and a degree
$d_2=9$ hypersurface. $E_2$ is the rank 3 bundle on $M_2$ defined by
$\{n_a\}=\{1,2,2,5\},\ m=10$. It is described by {\it exactly the same} \LG\
superpotential, but with $\Sigma^1\leftrightarrow \Lambda^4$,
$\Sigma^2\leftrightarrow \Lambda^2$.

As \LG\ theories, $(M_1,E_1)$ and $(M_2,E_2)$ are {\it isomorphic}. They have
exactly the same spectrum of states, and give rise to identical string
theories. Geometrically, however, they are quite different. Following \GVW, we
can calculate the orbifold Euler characteristics of $M_{1,2}$. They are
different; $\chi(M_1)=-104$, and $\chi(M_2)=-108$. Here we have (0,2)
theories on different manifolds, with
{\it different Euler characteristics}, which
are nonetheless isomorphic as \LG\ theories\foot{The number of
generations, of course, is the same in these two theories, $\half
c_3(E_1)=\half c_3(E_2)$.}.

{}From our brief encounter with (0,2) theories, it is clear that
we have entered a rich
and largely unexplored territory. It is likely that even greater surprises
await us.

\bigbreak\bigskip\bigskip\centerline{{\bf Acknowledgments}}\nobreak
\frenchspacing{
We would like to thank B. Greene,  E. Witten, and especially D. Morrison
for helpful discussions on the subject of (0,2) theories
and for their comments on a draft of this manuscript. J. Distler would like to
thank the Aspen Center for Physics, where part of this work was completed.
This work was supported by
NSF grant PHY90-21984 and by the A.P. Sloan Foundation. }

\appendix{}{(0,2) Conventions}

Our (0,2) superspace has coordinates $(z,\bar z,\theta^+,\theta^-)$. The
spinor derivatives are
$$\bar D_\pm=\pd{}{\theta^\pm}+\theta ^\mp\partial_{\bar z}\quad.$$
A chiral superfield $\Phi$ has, as its lowest component a scalar $\phi$, and
satisfies the chiral constraint
$$\bar D_+ \Phi=0$$
In components, we have
$$\Phi=\phi+\theta^-\psi+\theta^-\theta^+\partial_{\bar z}\phi$$
A Fermi superfield $\Lambda$ satisfied the same chiral constraint $\bar D_+
\Lambda=0$, but has as its lowest component a left-moving fermion $\lambda$.
Its component expansion is
$$\Lambda=\lambda+\theta^-l+\theta^-\theta^+\partial_{\bar z}\lambda$$
where $l$ is an auxiliary field.

We also need to introduce the (0,2) gauge multiplets. These consist of a real
superfield $V$, whose lowest component is a scalar, and a superfield $\CA$,
whose lowest component is the left-moving component of the gauge field, $a$.
A super-gauge transformation is
$$V\to V-i(\chi-\bar\chi),\qquad \CA\to\CA-i(\chi+\bar\chi)$$
with $\chi$ a chiral superfield $\bar D_+\chi=\bar D_-\bar \chi=0$.
In Wess-Zumino gauge, the nonzero components are
$$\eqalign{V&=\theta^-\theta^+\bar a\cr
\CA&=a+\theta^+\alpha-\theta^-\tilde\alpha+\theta^-\theta^+ D\cr}$$
where $\bar a$ is the right-moving component of the gauge field,
$\alpha,\tilde\alpha$ are left-moving gauginos, and $D$ is an auxiliary field.

Under a super-gauge transformation,
$$\Phi\to\ex{2iQ\chi}\Phi,\qquad \bar\Phi\to\ex{-2iQ\bar\chi}\bar\Phi,$$
so a gauge-invariant kinetic term for $\Phi$ is
$$\eqalign{S_\Phi&=\half\int\dd^2z\dd^2\theta\ (\partial-Q\CA)\bar{\tilde\Phi}
\tilde\Phi- \bar{\tilde\Phi}(\partial+Q\CA)\tilde\Phi\cr
&=\int\dd^2z\ (\partial-Qa)\bar\phi(\bar\partial+Q\bar
a)\phi+(\bar\partial-Q\bar a)\bar\phi(\partial+Qa)\phi
+\bar\psi(\partial+Qa)\psi\cr
&\qquad+Q(\tilde\alpha\bar\psi-\alpha\psi\bar\phi)-Q\bar\phi\phi D\cr}$$
where
$$\tilde\Phi=\ex{QV}\Phi,\qquad \bar{\tilde\Phi}= \ex{QV}\bar\Phi\quad.$$
An invariant kinetic term for $\Lambda$ is
$$\eqalign{S_\Lambda&=\half\int\dd^2z\dd^2\theta\ \bar\Lambda\ex{2QV}\Lambda\cr
&=\int\dd^2z\ \bar\lambda(\partial_{\bar z}+Q\bar a)\lambda +\half  \bar l
l\cr}$$

If we define the gauge-covariant spinor derivatives
$$\bar \CD_\pm=\ex{\pm V}\bar D_\pm\ex{\mp V},\qquad \CD=\partial+\CA,$$
we can write the gauge-invariant field strengths
$$\eqalign{\CF
&=[\CD,\bar\CD_+]=-\alpha+\theta^-(f+D)-\theta^-\theta^+\partial_{\bar
z}\alpha\cr
\bar\CF&=[\CD,\bar\CD_-]
=\tilde\alpha+\theta^+(f-D)-\theta^-\theta^+\partial_{\bar
z}\tilde\alpha\cr}$$
where $f=\partial_z\bar a-\partial_{\bar z}a$.
The gauge kinetic term is
$$\eqalign{S_\CF&=-{1\over 2e^2}\int \dd^2z\dd^2\theta\ \CF\bar\CF\cr
&={1\over e^2}\int \dd^2z\ (-\half f^2+\half D^2-\alpha\partial_{\bar
z}\tilde\alpha)\quad.\cr}$$

The Fayet-Iliopoulos D-term and the $U(1)$ $\theta$-term can be written as
$$\eqalign{S_D&={t\over 2}\int\dd^2z\dd\theta^-\CF-{\bar t\over
2}\int\dd^2z\dd\theta^+\bar\CF\cr
&=r\int \dd^2z\ D +i\theta\int \dd^2 z f\cr}$$
where $t=r+i\theta$.

Finally, a (0,2) superpotential has the form
$$\eqalign{S_\CW&=\int\dd^2z\dd\theta^- \Lambda
F(\Phi)+\int\dd^2z\dd\theta^+\overline\Lambda\ \overline{F(\Phi)}\cr
&=\int \dd^2z\ (l F(\phi)-\lambda\pd{F}{\phi}\psi)+ {\rm h.c.}\cr}$$

\listrefs
\end